\documentclass[pra,aps,reprint,superbib,superscriptaddress,twocolumn]{revtex4-1}
\usepackage{amsmath,bm,bbm}
\usepackage{amstext}
\usepackage{epsfig}
\usepackage{xcolor}
\usepackage{subfig}
\usepackage{graphicx}
\usepackage{multirow}
\usepackage{array}
\usepackage{tikz,pgfplots}
\usepackage{MnSymbol}
\usepackage{dsfont}
\usepackage{bbold}
\usepackage{physics}
\usepackage{mathtools}
\usepackage{hyperref}
\hypersetup{
    colorlinks=true,
    citecolor=gray,
    filecolor=blue,
    linkcolor=blue,
    urlcolor=red
}

\usepackage{mathrsfs}

\definecolor{dgreen}{rgb}{0,.5,0}
\definecolor{dred}{rgb}{.7,.0,.0}

\def\ddroit{{d}}
\newcommand{\be}{\begin{eqnarray}}
\newcommand{\ee}{\end{eqnarray}}

\newcommand{\bruno}[1]{{\textcolor{blue}{Bruno: #1 }} }
\newcommand{\manu}[1]{{\textcolor{dgreen}{ Manu: #1 }} }

\usepackage{xr}

\DeclarePairedDelimiter\ceil{\lceil}{\rceil}
\DeclarePairedDelimiter\floor{\lfloor}{\rfloor}
\newcommand{\br}{\mathbf{r}}
\newcommand{\dr}{d\mathbf{r}}
\newcommand{\tra}[1]{{\rm Tr}\left[#1\right]}
\newcommand{\blue}[1]{{\textcolor{blue}{ #1 }} }
\begin{document}

\title{
$N$-centered ensemble density-functional theory for open
systems
}
\author{Bruno Senjean}
\thanks{Corresponding author}
\email{bsenjean@gmail.com}
\affiliation{Instituut-Lorentz, Universiteit Leiden, P.O. Box 9506, 2300 RA Leiden, The Netherlands}
\affiliation{Division of Theoretical Chemistry, Vrije Universiteit Amsterdam, De Boelelaan 1083, 1081 HV Amsterdam, The Netherlands}
\author{Emmanuel Fromager}
\affiliation{Laboratoire de Chimie Quantique,
Institut de Chimie, CNRS/Universit\'e de Strasbourg,
4 rue Blaise Pascal, 67000 Strasbourg, France}


\begin{abstract}
Two (so-called left and right) variants of $N$-centered ensemble
density-functional theory (DFT) [Senjean and Fromager, Phys. Rev. A 98, 022513 (2018)] are presented. Unlike the original
formulation of the theory, these variants allow for the description of 
systems with a fractional electron number. While conventional DFT for open systems uses only the true electron
density as basic variable, left/right $N$-centered ensemble DFT relies instead
on (i) a fictitious ensemble density that integrates to a central
(integral) number $N$ of electrons, {\it and} (ii) a grand canonical
ensemble weight $\alpha$
which is equal to the deviation of the true electron number from $N$. 
Within such a formalism, the infamous derivative
discontinuity that appears when crossing an integral number of electrons
is described exactly through the dependence in $\alpha$ of the left and right 
$N$-centered ensemble Hartree-exchange-correlation 
density functionals. Incorporating $N$-centered ensembles into existing
density-functional embedding theories is expected to pave the way
towards the in-principle-exact
description of an open fragment by means of a pure-state $N$-electron many-body
wavefunction. Work is currently in progress in this direction.   
\end{abstract}

\maketitle

\section{Introduction}

Density-functional theory (DFT) has become over the last two decades the
method of choice for performing routine large-scale electronic structure calculations. This success is essentially
due to the relatively low computational cost of the method and its (often but not
always) good accuracy. In most applications, DFT is applied to closed
electronic  
systems, i.e. systems with an integral number $N$ of electrons.
However, at the formal level, there is no such a restriction.
In other words, DFT is in principle able to describe also systems with a
fractional electron number, as shown in the pioneering work of Perdew, Parr, Levy and
Balduz (PPLB)~\cite{perdew1982density}. The extension of DFT to open systems plays a
crucial role in the description of charged electronic
excitations~\cite{perdew1983physical}. More
recently, it became a key ingredient in the derivation of
DFT-based
embedding approaches such as 
partition
DFT~\cite{cohen2006hardness,cohen2007foundations,
elliott2009density,elliott2010partition,nafziger2011molecular,
tang2012fragment,mosquera2013partition,niffenegger2019density},
potential-functional embedding theory~\cite{huang2011potential},
and
frozen density embedding theory for non-integer subsystems'
particle numbers~\cite{fabiano2014frozen}.\\

The density $n(\br)$ of an open system integrates to a fractional number
$\mathcal{N}$ of electrons. As shown in Refs.~\cite{perdew1982density,perdew1983physical}, the ground-state
energy of such a system varies linearly with the deviation
$\alpha=\mathcal{N}-\floor{\mathcal{N}}$ of the true electron number
from its floor integral value. In the language of ensemble
DFT~\cite{elliott2010partition,kraisler2013piecewise},
the density is nothing but a weighted sum of ground-state densities
$n^{N-1}(\br)$ and $n^N(\br)$ integrating to $N-1=\floor{\mathcal{N}}$
and $N=\ceil{\mathcal{N}}$ electrons, respectively: 
\be\label{eq:open_sys_dens_as_ens_dens}
n(\br)=(1-\alpha)n^{N-1}(\br)+\alpha\, n^N(\br),
\ee
where, as readily seen, the ensemble weights are $\alpha$ and
$(1-\alpha)$, respectively. As shown in
Ref.~\cite{kraisler2013piecewise}, the explicit expression in
Eq.~(\ref{eq:open_sys_dens_as_ens_dens}) is convenient for constructing
density functional approximations to the open-system
exchange-correlation (xc) energy $E_{\rm xc}[n]$ as functions of
$\alpha$.\\

Quite recently, the authors have proposed an in-principle-exact reformulation of the
fundamental gap problem in DFT~\cite{senjean2018unified}. For that purpose, they introduced the
concept of $N$-centered ensemble where, unlike in the true physical
ensemble described in Eq.~(\ref{eq:open_sys_dens_as_ens_dens}), the
ensemble density integrates to an {\it integral} number of electrons.
The practical advantage of such a reformulation is that the infamous
derivative discontinuity contribution to the true gap~\cite{perdew1983physical} can be expressed
as an ensemble Hartree (H) xc energy derivative with respect to the
ensemble weight~\cite{senjean2018unified}, exactly like in DFT for
canonical ensembles~\cite{gross1988density}. In the original formulation
of the theory~\cite{senjean2018unified}, the ensemble weights have no physical
meaning. They are just auxiliary variables which, through their
variations,
enable the extraction of quantities of interest like the ionization
potential (IP), the electron affinity (EA) or, when a single weight is
used, the fundamental gap. In fact, $N$-centered ensemble DFT allows for a direct
extraction of individual energies for the neutral,
anionic and cationic systems~\cite{senjean2018unified}, which are all closed systems, in a single calculation. In this
work, we show how the theory can be adapted to open
systems. By considering ionization and affinity processes separately, we
obtain two variants (referred to as {\it left} and {\it right}) of
$N$-centered ensemble DFT where the ensemble weight is now connected to
$\alpha$ [see Eq.~(\ref{eq:open_sys_dens_as_ens_dens})].\\

The paper is organized as follows. After a brief review on the original
formulation of $N$-centered ensemble DFT in
Sec.~\ref{subsec:original-Nc-eDFT}, the concept of left and right
$N$-centered ensembles is introduced (Sec.
\ref{subsec:L-R-Nc-ensembles}). The left/right variants of
$N$-centered ensemble DFT are then derived in 
Sec.~\ref{subsec:L-R-Nc-eDFT}. In this context we obtain an in-principle-exact
reformulation of the IP/EA theorem, as discussed in
Sec.~\ref{subsec:IP-EA_th}. An explicit connection (through scaling
relations) between left and right
ensemble density functionals is then made in
Sec.~\ref{subsec:left_or_right}. Finally, a brief discussion on density-driven
correlation
effects, whose importance 
in canonical
ensembles
was recently revealed~\cite{gould2019density,gould2020density,fromager2020individual},
is proposed in Sec.~\ref{subsec:DDeffects}. As a conclusion to the ``Theory" Sec.~\ref{sec:theory}, we
compare left/right $N$-centered ensemble DFT with conventional PPLB-DFT
for open systems
(Sec.~\ref{subsec:comparison_with_Levy-Perdew-approach}).
A proof of concept study of the Hubbard dimer model is presented in
Sec.~\ref{sec:hubbard}, with a particular emphasis on the weight-dependence of the
Hxc left/right ensemble functionals. Conclusions and perspectives are
finally given in Sec.~\ref{sec:conclusions}.      
\section{Theory}\label{sec:theory}

\subsection{$N$-centered ensembles }\label{subsec:original-Nc-eDFT}

In a previous work~\cite{senjean2018unified}, we introduced the concept
of $N$-centered ensemble for the purpose of calculating fundamental
gaps, in principle exactly, within DFT. In standard approaches, the
number of electrons is considered to vary continuously between two integers, thus
allowing for the description of ionization or affinity
processes~\cite{kraisler2013piecewise,baerends2017kohn,
andrade2011prediction,atalla2016enforcing,
bhandari2018fundamental,elmaslmane2018accuracy,
imamura2011linearity,kronik2012excitation,zheng2011improving,
zheng2013nonempirical,li2015local,stein2012curvature,
hait2018delocalization,chan1999fresh,cohen2008fractional,
gould2014kohn,hodgson2017interatomic,benitez2016kohn,
gorling2015exchange,li2017extending,
gould2019what,hellgren2019strong,kraisler2019discontinuous}. The
situation is completely different in an $N$-centered ensemble where, by
construction, the number of electrons (which is obtained by integration of
the $N$-centered ensemble density) is not affected by the charged
excitation. It remains equal to the so-called central number $N$ of
electrons which is an {\it integer}. Formally, an $N$-centered ensemble is
described by the following density matrix operator~\cite{senjean2018unified}:    
\be\label{eq:original_N-centered_ens_twoweights}
\hat{\Gamma}^{\left\{N,\xi_-,\xi_+\right\}}&=&\xi_-\hat{\Gamma}^{N-1}+\xi_+\hat{\Gamma}^{N+1}
\nonumber\\
&&+\left(1-\dfrac{(N-1)}{N}\xi_- -
\dfrac{(N+1)}{N}\xi_+\right)\hat{\Gamma}^{N},
\ee
where $\xi_-$ and $\xi_+$ are ensemble weights assigned to the cationic
$(N-1)$-electron and anionic $(N+1)$-electron systems, respectively.
Each individual density matrix operator $\hat{\Gamma}^{N_e}$, where
$N_e=N,N\pm1$, is used to describe an $N_e$-electron ground state. If
the latter can be described with a pure-state wavefunction $\Psi^{N_e}$ then
$\hat{\Gamma}^{N_e}\equiv\ket{\Psi^{N_e}}\bra{\Psi^{N_e}}$. In case of
degeneracy, it
may be written as a convex combination of pure $N_e$-electron states
$\hat{\Gamma}^{N_e}\equiv\sum_I\lambda_I\ket{\Psi_I^{N_e}}\bra{\Psi_I^{N_e}}$,
where $\sum_I\lambda_I=1$. As readily seen from
Eq.~(\ref{eq:original_N-centered_ens_twoweights}), the
number of electrons within the $N$-centered ensemble is
$\tra{\hat{\Gamma}^{\left\{N,\xi_-,\xi_+\right\}}\hat{\mathcal{N}}}=N$
where $\tra{\ldots}$ denotes the trace, $\hat{\mathcal{N}}=\int
\dr\;\hat{n}(\br)$ is the electron counting operator, and $\hat{n}(\br)$
is the density operator at position $\br$.\\

As shown in Ref.~\cite{senjean2018unified}, using two independent
weights $\xi_-$ and $\xi_+$ allows for a separate extraction of the IP 
and EA from the $N$-centered ensemble energy
$\tra{\hat{\Gamma}^{\left\{N,\xi_-,\xi_+\right\}}\hat{H}}$, where
$\hat{H}$ is the (second-quantized) Hamiltonian. In the particular case where
$\xi_-=\xi_+=\xi$, the $N$-centered ensemble density matrix operator
simplifies as follows:  
\be
\hat{\Gamma}^{\left\{N,\xi\right\}}=\xi\hat{\Gamma}^{N-1}+\xi\hat{\Gamma}^{N+1}
+(1-2\xi)\hat{\Gamma}^{N},
\ee
so that the fundamental gap can be extracted directly by 
differentiating the corresponding $N$-centered ensemble energy
$\tra{\hat{\Gamma}^{\left\{N,\xi\right\}}\hat{H}}$ with
respect to $\xi$~\cite{senjean2018unified}. The potential advantage of such a
formalism over conventional approaches is that the infamous
derivative discontinuity contribution to the gap can be written as the
derivative of the $N$-centered ensemble Hxc density functional with respect to the
ensemble weight $\xi$, exactly
like in DFT for canonical
ensembles~\cite{gross1988density,levy1995excitation}.\\

Let us stress that, in the current
formulation of $N$-centered ensemble DFT, the ensemble weights have no
physical meaning. They are just convenient auxiliary variables.
Actually, the properties of interest [namely the ground-state
energies of the neutral ($N$-electron), cationic and anionic systems]
should not depend on the value of the ensemble weights.   
In the rest of this work, we explore two variants of the
theory which are directly applicable to 
open
systems. For such systems, a convenient and physical choice for the
ensemble weight value is the deviation of the
(fractional) electron number $\mathcal{N}$ from either its integral
floor value 
$\floor{\mathcal{N}}$ or the ceiling one $\ceil{\mathcal{N}}$.

\subsection{Left and right $N$-centered
ensembles}\label{subsec:L-R-Nc-ensembles}

Starting from Eq.~(\ref{eq:original_N-centered_ens_twoweights}), we
explore in this section different choices of $N$-centered ensemble weights $\xi_-$
and $\xi_+$ for the
purpose of describing an open $\mathcal{N}$-electron system. In order to
treat both $\ceil{\mathcal{N}}=N$ and $\floor{\mathcal{N}}=N$ scenarios,
thus allowing for the description of fluctuations around the central
(integral) number $N$ of electrons, we introduce what we will refer to as {\it left} (subscript `$-$') and
{\it right} (subscript `$+$')
$N$-centered ensemble density matrix operators, respectively:
\be\label{eq:definitive_def_L-R_ens}
\hat{\Gamma}_-^{\{N,\alpha\}}=(1-\alpha)\hat{\Gamma}^{N}+\dfrac{N\alpha}{N - 1}\hat{\Gamma}^{N - 1},
\ee
\be\label{eq:definitive_def_L-R_ens_right}
\hat{\Gamma}_+^{\{N,\alpha\}}=(1-\alpha)\hat{\Gamma}^{N}+\dfrac{N\alpha}{N
+ 1}\hat{\Gamma}^{N + 1},
\ee
where $0\leq\alpha\leq1$. Note that, by construction, the number of
electrons within these two 
ensembles is the central number $N$:  
\be
\tra{\hat{\Gamma}_-^{\{N,\alpha\}}\hat{\mathcal{N}}}=\tra{\hat{\Gamma}_+^{\{N,\alpha\}}\hat{\mathcal{N}}}=N.
\ee 
Moreover, left and right ensembles can be 
connected to each other as follows:
\begin{eqnarray}\label{eq:link_left_right_ensembles}
\dfrac{N}{N-1}\hat{\Gamma}^{\{N-1,\alpha\}}_+
= \hat{\Gamma}^{\{N,1-\alpha\}}_-.
\end{eqnarray}
These ensembles are just special
cases of the original   
$N$-centered one introduced in Ref.~\cite{senjean2018unified}. Indeed, with the
notations of Eq.~(\ref{eq:original_N-centered_ens_twoweights}), we have     
\be\label{eq:left_from_two_weight}
\hat{\Gamma}_-^{\{N,\alpha\}}=
\hat{\Gamma}^{\left\{N,\xi_-=\frac{N\alpha}{N-1},\xi_+=0\right\}},
\ee
and
\be\label{eq:right_from_two_weight}
\hat{\Gamma}_+^{\{N,\alpha\}}=\hat{\Gamma}^{\left\{N,\xi_-=0,\xi_+=\frac{N\alpha}{N+1}\right\}}.
\ee
The original (two-weight) $N$-centered ensemble can actually be recovered from the left and
right ones as follows:
\be\label{eq:connection_L-R_two-weight_ens}
\hat{\Gamma}^{\left\{N,\xi_-,\xi_+\right\}}&=&
\dfrac{(N-1)}{2N}\hat{\Gamma}_-^{\{N,2\xi_-\}}+\dfrac{(N+1)}{2N}\hat{\Gamma}_+^{\{N,2\xi_+\}}.
\ee

\begin{figure}
\centering
\resizebox{\columnwidth}{!}{
\includegraphics[scale=1]{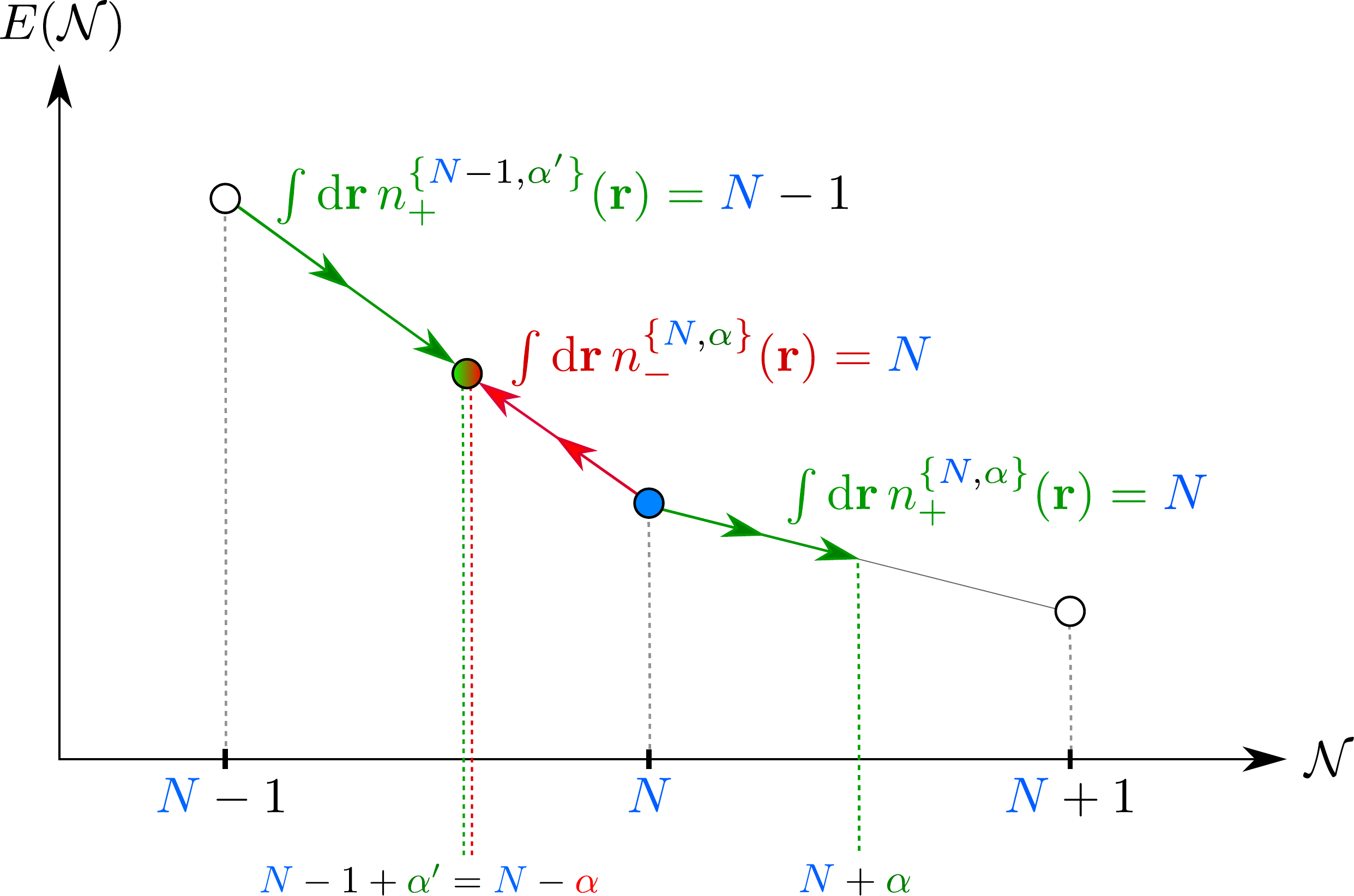}
}
\caption{Graphical illustration of left and right $N$-centered
ensembles. The shorthand notation
$n^{\left\{N,\alpha\right\}}_{\pm}(\br)\equiv
n_{\hat{\Gamma}_\pm^{\{N,\alpha\}}}(\br)$ is used.
}
\label{fig:l-r-schematics}
\end{figure}

Following the standard description of open systems in DFT~\cite{perdew1982density}, we
propose to use for the value of $\alpha$ in
Eqs.~(\ref{eq:definitive_def_L-R_ens}) and (\ref{eq:definitive_def_L-R_ens_right}) the deviation of the true (fractional)
electron number $\mathcal{N}$, that we assume to vary in the range
$N-1<\mathcal{N}<N+1$, from the central number $N$, i.e.
\be
\alpha=\abs{\mathcal{N}-N}.
\ee
If $\ceil{\mathcal{N}}=N$, then $\alpha=N-\mathcal{N}$ and the open
system will be described by the left $N$-centered ensemble of
Eq.~(\ref{eq:definitive_def_L-R_ens}). If
$\floor{\mathcal{N}}=N$, then $\alpha=\mathcal{N}-N$ and 
the right $N$-centered ensemble of
Eq.~(\ref{eq:definitive_def_L-R_ens_right}) will be used instead. This
is illustrated graphically in Fig.~\ref{fig:l-r-schematics}.\\
 
Let us stress that these ensembles are {\it not}
the physical ones, which are described by the
following density matrix operators~\cite{kraisler2013piecewise}:
\be\label{eq:gamma_matchalN}
\hat{\Gamma}\left(\mathcal{N}\right)
&\overset{\ceil{\mathcal{N}}=N}{\equiv}&
 \hat{\Gamma}\left(N-\alpha\right)
=\left(1-\alpha\right)\hat{\Gamma}^{N}+\alpha \hat{\Gamma}^{N-1},
\nonumber\\
\hat{\Gamma}\left(\mathcal{N}\right)
&\overset{\floor{\mathcal{N}}=N}{\equiv}&
 \hat{\Gamma}\left(N+\alpha\right)
=\left(1-\alpha\right)\hat{\Gamma}^{N}+\alpha \hat{\Gamma}^{N+1}.
\ee
Nevertheless, they can be used as auxiliary ensembles
from which the exact properties of the true physical open system can be
extracted. Indeed, according to Eqs.~(\ref{eq:definitive_def_L-R_ens})
and (\ref{eq:definitive_def_L-R_ens_right}),   
\be\label{eq:GammaN_from_L-R_ens}
\hat{\Gamma}^{N}&=&\hat{\Gamma}_\pm^{\{N,\alpha=0\}}=
\hat{\Gamma}_\pm^{\{N,\alpha\}}-\alpha\dfrac{\partial \hat{\Gamma}_\pm^{\{N,\alpha\}}}{\partial
\alpha}
\ee
and
\be\label{eq:GammaNpm1_from_LorR_ens}
\hat{\Gamma}^{N\pm1}&=&\dfrac{N\pm1}{N}
\left(\dfrac{\partial \hat{\Gamma}_\pm^{\{N,       \alpha\}}}{\partial
\alpha}+\hat{\Gamma}^{N}\right)
\nonumber\\
&=&
\dfrac{N\pm1}{N}\left(\hat{\Gamma}_\pm^{\{N,\alpha\}}
+(1-\alpha)\dfrac{\partial \hat{\Gamma}_\pm^{\{N,       \alpha\}}}{\partial
\alpha}\right),
\ee 
thus leading to the final expressions
\be\label{eq:true_ens_from_left}
\hat{\Gamma}\left(\mathcal{N}\right)
&\overset{\ceil{\mathcal{N}}=N}{\equiv}&
 \hat{\Gamma}\left(N-\alpha\right)
\nonumber\\
&=&\left(1-\dfrac{\alpha}{N}\right)\hat{\Gamma}_-^{\{N,\alpha\}}-\dfrac{\alpha(1-\alpha)}{N}\dfrac{\partial
\hat{\Gamma}_-^{\{N,\alpha\}}}{\partial
\alpha},
\ee
and
\be\label{eq:true_ens_from_right}
\hat{\Gamma}\left(\mathcal{N}\right)
&\overset{\floor{\mathcal{N}}=N}{\equiv}&
 \hat{\Gamma}\left(N+\alpha\right)
\nonumber\\
&=&\left(1+\dfrac{\alpha}{N}\right)\hat{\Gamma}_+^{\{N,\alpha\}}+\dfrac{\alpha(1-\alpha)}{N}\dfrac{\partial
\hat{\Gamma}_+^{\{N,\alpha\}}}{\partial
\alpha}.
\ee
A direct consequence of Eqs.~(\ref{eq:true_ens_from_left}) and
(\ref{eq:true_ens_from_right}) is that the physical energy
$E\left(\mathcal{N}\right)=\tra{\hat{\Gamma}\left(\mathcal{N}\right)\hat{H}}$ of an open
system can be extracted from a left or right $N$-centered ensemble as
follows:
\be\label{eq:true_ener_from_left}
E\left(\mathcal{N}\right)
&\overset{\ceil{\mathcal{N}}=N}{\equiv}&
 E\left(N-\alpha\right)
\nonumber\\
&=&\left(1-\dfrac{\alpha}{N}\right)\mathscr{E}_-^{\{N,\alpha\}}-\dfrac{\alpha(1-  \alpha)}{N}\dfrac{\partial
\mathscr{E}_-^{\{N,\alpha\}}}{\partial
\alpha},
\ee
and
\be\label{eq:true_ener_from_right}
E\left(\mathcal{N}\right)
&\overset{\floor{\mathcal{N}}=N}{\equiv}&
 E\left(N+\alpha\right)
\nonumber\\
&=&\left(1+\dfrac{\alpha}{N}\right)\mathscr{E}_+^{\{N,\alpha\}}+\dfrac{\alpha(1-  \alpha)}{N}\dfrac{\partial
\mathscr{E}_+^{\{N,\alpha\}}}{\partial
\alpha},
\ee
where
$\mathscr{E}_\pm^{\{N,\alpha\}}=\tra{\hat{\Gamma}_\pm^{\{N,\alpha\}}\hat{H}}$
are the left/right $N$-centered ensemble energies.\\

Similarly, the
true density
$n_{\hat{\Gamma}\left(\mathcal{N}\right)}(\br)=\tra{\hat{\Gamma}\left(\mathcal{N}\right)\hat{n}(\br)}$
can be obtained as follows from the left or right $N$-centered ensemble
ones
$n_{\hat{\Gamma}_\pm^{\{N,\alpha\}}}(\br)=\tra{\hat{\Gamma}_\pm^{\{N,\alpha\}}\hat{n}(\br)}$:
\be\label{eq:true_dens_from_left}
&&n_{\hat{\Gamma}\left(\mathcal{N}\right)}(\br)
\overset{\ceil{\mathcal{N}}=N}{\equiv}
n_{\hat{\Gamma}\left(N-\alpha\right)}(\br)
\nonumber\\
&&=\left(1-\dfrac{\alpha}{N}\right)n_{\hat{\Gamma}_-^{\{N,\alpha\}}}(\br)-\dfrac{\alpha(1-
\alpha)}{N}\dfrac{\partial n_{\hat{\Gamma}_-^{\{N,\alpha\}}}(\br)}{\partial
\alpha},
\ee
and
\be\label{eq:true_dens_from_right}
&&n_{\hat{\Gamma}\left(\mathcal{N}\right)}(\br)
\overset{\floor{\mathcal{N}}=N}{\equiv}
n_{\hat{\Gamma}\left(N+\alpha\right)}(\br)
\nonumber\\
&&=\left(1+\dfrac{\alpha}{N}\right)n_{\hat{\Gamma}_+^{\{N,\alpha\}}}(\br)+\dfrac{\alpha(1-  \alpha)}{N}\dfrac{\partial
n_{\hat{\Gamma}_+^{\{N,\alpha\}}}(\br)}{\partial
\alpha}.
\ee

Let us finally point out that, according to Eqs.~(\ref{eq:GammaN_from_L-R_ens}) and (\ref{eq:GammaNpm1_from_LorR_ens}), 
\be
\hat{\Gamma}^{N-1}-\hat{\Gamma}^{N}=-\dfrac{1}{N}
\left(\hat{\Gamma}_-^{\{N,\alpha\}}
+(1-\alpha-N)\dfrac{\partial\hat{\Gamma}_-^{\{N,\alpha\}}}{\partial
\alpha}
\right),
\ee
and
\be
\hat{\Gamma}^{N}-\hat{\Gamma}^{N+1}=-\dfrac{1}{N}
\left(\hat{\Gamma}_+^{\{N,\alpha\}}
+(1-\alpha+N)\dfrac{\partial\hat{\Gamma}_+^{\{N,\alpha\}}}{\partial
\alpha}
\right).
\ee
Consequently, the IP and EA can be extracted from the left and right
$N$-centered ensemble energies as follows:
\be\label{eq:IP_from_n-centered_ens}
I^N&=&\tra{\left(\hat{\Gamma}^{N-1}-\hat{\Gamma}^{N}\right)\hat{H}}
\nonumber\\
&=&-\dfrac{1}{N}
\left(\mathscr{E}_-^{\{N,\alpha\}}
+(1-\alpha-N)\dfrac{\partial \mathscr{E}_-^{\{N,\alpha\}}}{\partial
\alpha}
\right),
\ee
and
\be\label{eq:EA_from_n-centered_ens}
A^N&=&\tra{\left(\hat{\Gamma}^{N}-\hat{\Gamma}^{N+1}\right)\hat{H}}
\nonumber\\
&=&-\dfrac{1}{N}
\left(\mathscr{E}_+^{\{N,\alpha\}}
+(1-\alpha+N)\dfrac{\partial \mathscr{E}_+^{\{N,\alpha\}}}{\partial
\alpha}
\right).
\ee

As shown in the following, by deriving a DFT for left and right
$N$-centered ensembles, we obtain from Eqs.~(\ref{eq:true_ener_from_left}), (\ref{eq:true_ener_from_right}),
(\ref{eq:true_dens_from_left}), and (\ref{eq:true_dens_from_right}) a
novel and in-principle-exact density-functional description of
open systems. 

\subsection{Left/right $N$-centered ensemble density-functional theory}\label{subsec:L-R-Nc-eDFT}

In this section we derive a variational Kohn--Sham (KS) DFT expression
for the left/right $N$-centered ensemble energies.
As a result, the open system problem
will be mapped [{\it via} Eqs.~(\ref{eq:true_ener_from_left}) and
(\ref{eq:true_ener_from_right})] onto an auxiliary $N$-electron
non-interacting ensemble, where the (central) number $N$ is an {\it
integer}. The derivation follows closely the one
presented in Ref.~\cite{senjean2018unified} for two-weight $N$-centered ensembles [see
Eq.~(\ref{eq:original_N-centered_ens_twoweights})].\\
 
Let us consider the usual {\it ab-initio} quantum
chemical Hamiltonian $\hat{H}=\hat{T}+\hat{W}_{\rm ee}+\hat{V}_{\rm ext}$ where $\hat{T}$ and $\hat{W}_{\rm ee}$ are the
kinetic energy and two-electron repulsion operators, respectively, and
$\hat{V}_{\rm ext}=\int
d{\br}\,v_{\rm ext}
(\br)\,\hat{n}(\br)$ is the external potential operator. For an isolated
molecule, the external local potential $v_{\rm ext}(\br)$ is simply the nuclear potential. If we introduce the analog for
left/right $N$-centered ensembles of Levy's constrained-search universal
functional, 
\be\label{eq:LL_L-R_fun}
F_\pm^{\{N,\alpha\}}[n]&=&
\min_{\hat{\gamma}_\pm^{\{N,\alpha\}}\rightarrow n}\left\{
\tra{\hat{\gamma}_\pm^{\{N,\alpha\}}\left(\hat{T}+\hat{W}_{\rm ee}\right)}
\right\}
\nonumber\\
&=&\tra{\hat{\Gamma}_\pm^{\{N,\alpha\}}[n]\left(\hat{T}+\hat{W}_{\rm          ee}\right)},
\ee
where the minimization is performed over left/right $N$-centered
ensemble density matrix operators 
$\hat{\gamma}_\pm^{\{N,\alpha\}}$ 
that fulfill the density constraint
\be\label{eq:N-centered_dens_constraint}
n_{\hat{\gamma}_\pm^{\{N,\alpha\}}}(\mathbf{r})=\tra{\hat{\gamma}_\pm^{\{N,     \alpha\}}\hat{n}(\br)}=n(\mathbf{r})
,
\ee
then 
it becomes possible to express the exact left/right $N$-centered
ensemble energies variationally as follows:
\be\label{eq:var_princ_L-R_dens}
\mathscr{E}_\pm^{\{N,\alpha\}}=\min_{n\rightarrow
N}\left\{F_\pm^{\{N,\alpha\}}[n]+
\int\ddroit{\mathbf{r}}\,v_{\rm ext}(\mathbf{r})n(\mathbf{r})\right\}.
\ee
The minimum is reached when the density equals the exact left/right $N$-centered
ensemble one $n_{\hat{\Gamma}_\pm^{\{N,\alpha\}}}$.
Note that the true (i.e. interacting) density matrix operators $\hat{\Gamma}_\pm^{\{N,
\alpha\}}$ in Eqs.~(\ref{eq:definitive_def_L-R_ens}) and
(\ref{eq:definitive_def_L-R_ens_right}) would be obtained by solving the ground-state Schr\"{o}dinger equation
$\hat{H}\Psi^{N_{\rm e}}=E^{N_{\rm e}}\Psi^{N_{\rm e}}$ for $N_{\rm
e}=N$ and $N_{\rm
e}=N\pm1$ electrons. Like in regular DFT, we bypass this complicated task by
introducing the following KS decomposition, 
\be\label{eq:KS_decomp}
F_\pm^{\{N,\alpha\}}[n]=T_{\rm s\pm}^{\{N,\alpha\}}[n]+E_{{\rm
Hxc}\pm}^{\{N,\alpha\}}[n],
\ee
where the non-interacting density-functional kinetic energy contribution reads
\be\label{eq:Ts_L-R}
T_{\rm s\pm}^{\{N,\alpha\}}[n]&=&
\min_{\hat{\gamma}_\pm^{\{N,\alpha\}}\rightarrow n}\left\{
\tra{\hat{\gamma}_\pm^{\{N,\alpha\}}\hat{T}
}
\right\}
\nonumber\\
&=&\tra{\hat{\gamma}_{s\pm}^{\{N,\alpha\}}[n]\,\hat{T}
}
,
\ee
and $E_{{\rm Hxc}\pm}^{\{N,\alpha\}}[n]$ is the left/right $N$-centered
ensemble Hxc functional. Note that this
functional is, for a fixed density $n$, $\alpha$-dependent. It differs
from the conventional ($N$-electron)
ground-state Hxc functional $E_{\rm Hxc}[n]$, which is recovered when
$\alpha=0$:
\be\label{eq:alpha_zero_limit_Hxcfun}
E_{{\rm Hxc}\pm}^{\{N,\alpha=0\}}[n]=E_{\rm Hxc}[n].
\ee
As shown in the following, modeling the dependence in $\alpha$ of the
Hxc functional in this context is analogous to modeling the derivative discontinuity in
conventional DFT for open systems. In the light of 
Ref.~\cite{gould2017hartree}, we define the exact Hx contribution as
follows:
\be\label{eq:EEXX_l-r}
E_{{\rm
Hx}\pm}^{\{N,\alpha\}}[n]=\tra{\hat{\gamma}_{s\pm}^{\{N,\alpha\}}[n]\,\hat{W}_{\rm
ee}},
\ee
where $\hat{\gamma}_{s\pm}^{\{N,\alpha\}}[n]$ is the minimizing
left/right $N$-centered ensemble density matrix operator in
Eq.~(\ref{eq:Ts_L-R}). According to Eqs.~(\ref{eq:LL_L-R_fun}), (\ref{eq:KS_decomp}), and
(\ref{eq:Ts_L-R}), the correlation contribution can then be
expressed as
\be
E_{{\rm
c}\pm}^{\{N,\alpha\}}[n]&=&
\tra{\hat{\Gamma}_\pm^{\{N,\alpha\}}[n]\left(\hat{T}+
\hat{W}_{\rm ee}\right)}
\nonumber\\
&&-\tra{\hat{\gamma}_{s\pm}^{\{N,\alpha\}}[n]\left(\hat{T}+
\hat{W}_{\rm ee}\right)}
<0.
\ee
 
Let us now return to the variational ensemble energy expression of
Eq.~(\ref{eq:var_princ_L-R_dens}). By inserting the exact decomposition
of Eq.~(\ref{eq:KS_decomp}) we obtain, according to
Eq.~(\ref{eq:Ts_L-R}), the final 
expressions 
\be\label{eq:KS1}
\mathscr{E}_\pm^{\{N,\alpha\}}&=&
 \min_{\hat{\gamma}_\pm^{\{N,\alpha\}}}
\Big\{
{\rm
Tr}\left[
\hat{\gamma}_\pm^{\{N,\alpha\}}
\hat{h}\right]
+
E^{\{N,\alpha\}}_{\rm Hxc\pm}[n_{\hat{\gamma}_\pm^{\{N,\alpha\}}}]
\Big\}
\nonumber\\
&=&\tra{\hat{\gamma}_{\rm s\pm}^{\{N,\alpha\}}
\hat{h}}+E^{\{N,\alpha\}}_{\rm Hxc\pm}[n_{\hat{\gamma}_{{\rm s}\pm}^{\{N,\alpha\}}}],
\ee
where $\hat{h}=\hat{T}+\hat{V}_{\rm ext}$. Like in regular 
ensemble DFTs, the minimizing
(non-interacting) density matrix operators
$\hat{\gamma}_{\rm s\pm}^{\{N,\alpha\}}$
in Eq.~(\ref{eq:KS1}) reproduce the exact interacting
left/right $N$-centered ensemble densities,
\be
n_{\hat{\gamma}_{\rm s-}^{\{N,\alpha\}}}(\br) 
&=& \left(1+\dfrac{
\alpha}{N-1}\right)\sum^{N-1}_{i=1}\abs{\varphi^{\{N,\alpha\}}_{i-}(\mathbf{r})}^2
\nonumber\\
&&+ (1-\alpha)\abs{\varphi^{\{N,\alpha\}}_{
N-}(\mathbf{r})}^2,\nonumber
\\
&=& n_{\hat{\Gamma}_-^{\{N,\alpha\}}}(\br),
\ee
and
\be
n_{\hat{\gamma}_{\rm s+}^{\{N,\alpha\}}}(\br) 
&=& \left(1-\dfrac{
\alpha}{N+1}\right)\sum^{N}_{i=1}\abs{\varphi^{\{N,\alpha\}}_{i+}(\mathbf{r})}^2
\nonumber\\
&&+ \dfrac{N\alpha}{N+1}\abs{\varphi^{\{N,\alpha\}}_{
(N+1)+}(\mathbf{r})}^2,\nonumber
\\
&=& n_{\hat{\Gamma}_+^{\{N,\alpha\}}}(\br).
\ee
The KS orbitals are obtained by
solving the following self-consistent equations,
\be\label{eq:sc-KS-L-R_eqs}
\left( -\dfrac{\nabla^2}{2} + v^{\{N,\alpha\}}_{\rm s\pm}(\br)\right) 
\varphi^{\{N,\alpha\}}_{i\pm}(\mathbf{r})=
\varepsilon_{i\pm}^{\{N,\alpha\}}
\varphi^{\{N,\alpha\}}_{i\pm}(\mathbf{r}),
\ee 
where the KS potential reads 
\be\label{eq:KS_pot_l-R}
v^{\{N,\alpha\}}_{\rm s\pm}(\br)= v_{\rm ext}(\mathbf{r})+
v^{\{N,\alpha\}}_{\rm Hxc\pm}[n_{\hat{\gamma}_{\rm s\pm}^{\{N,\alpha\}}}](\br)
,
\ee
and 
\be\label{eq:Hxc_pot_d_over_dn}
v^{\{N,\alpha\}}_{\rm Hxc\pm}[n](\br)=
\dfrac{\delta E^{\{N,\alpha\}}_{\rm
Hxc\pm}[n]}{\delta n(\br)}.
\ee
The final step in the formulation of left/right $N$-centered ensemble
DFT consists in connecting the true energy of the (interacting) open system under
study to the KS orbital energies.
For convenience, we use the Levy--Zahariev (LZ)
shift-in-potential
procedure~\cite{levy2014ground,senjean2018unified,deur2019ground},
\be
v^{\{N,\alpha\}}_{\rm
Hxc\pm}[n](\br)\rightarrow\overline{v}^{\{N,\alpha\}}_{\rm
Hxc\pm}[n](\br)
,
\ee
where
\be\label{eq:LZ_shift_L-R}
\overline{v}^{\{N,\alpha\}}_{\rm
Hxc\pm}[n](\br)&=&
v^{\{N,\alpha\}}_{\rm
Hxc\pm}[n](\br)
\nonumber\\
&&+
\dfrac{E^{\{N,\alpha\}}_{\rm
Hxc\pm}[n]-\int d\br \,v^{\{N,\alpha\}}_{\rm
Hxc\pm}[n](\br)n(\br)}{\int d\br\, n(\br)}.
\nonumber\\
\ee
Note that 
like in the original version of $N$-centered ensemble
DFT~\cite{senjean2018unified}, since 
\be
E^{\{N,\alpha\}}_{\rm
Hxc\pm}[n]=\int d\br \,\overline{v}^{\{N,\alpha\}}_{\rm
Hxc\pm}[n](\br)\,n(\br),
\ee
the total left/right $N$-centered ensemble KS energies 
match the interacting ones once the LZ shift
$\varepsilon_{i\pm}^{\{N,\alpha\}}\rightarrow\overline{\varepsilon}_{i\pm}^{\{N,\alpha\}}$
has been applied to the orbital energies
[see
Eqs.~(\ref{eq:sc-KS-L-R_eqs}) and (\ref{eq:KS_pot_l-R})], i.e.
\be\label{eq:N-centered_ener_DFT_left}
\mathscr{E}_-^{\{N,\alpha\}}
&= & 
\left(1+\dfrac{
\alpha}{N-1}\right)\sum^{N-1}_{i=1}\overline{\varepsilon}_{i-}^{\{N,\alpha\}}
+ (1-\alpha)\overline{\varepsilon}_{N-}^{\{N,\alpha\}}
,
\ee
and
\be\label{eq:N-centered_ener_DFT_right}
\mathscr{E}_+^{\{N,\alpha\}}&=& \left(1-\dfrac{
\alpha}{N+1}\right)\sum^{N}_{i=1}\overline{\varepsilon}_{i+}^{\{N,     \alpha\}}
+ \dfrac{N\alpha}{N+1}\overline{\varepsilon}_{(N+1)+}^{\{N,\alpha\}}.
\ee
Moreover, by applying the Hellmann--Feynman theorem to the variational
energy expressions in Eq.~(\ref{eq:KS1}), it comes 
\be\label{eq:deriv_N-centered_ener_DFT_left}
&&\dfrac{\partial
\mathscr{E}_-^{\{N,\alpha\}}}{\partial\alpha}
-\left.\dfrac{\partial E^{\{N,\alpha\}}_{\rm
Hxc-}[n]}{\partial \alpha}\right|_{n=
n_{\hat{\gamma}_{\rm s-}^{\{N,\alpha\}}}}
\nonumber\\
&=&
\dfrac{N}{N-1}\sum^{N-1}_{i=1}{\varepsilon}_{i-}^{\{N,\alpha\}}
-\sum^{N}_{i=1}{\varepsilon}_{i-}^{\{N,\alpha\}}
\nonumber\\
&=&
\dfrac{1}{N-1}\sum^{N-1}_{i=1}\Bigg({\varepsilon}_{i-}^{\{N,\alpha\}}-{\varepsilon}_{N-}^{\{N,\alpha\}}\Bigg)
\nonumber\\
&=&
\dfrac{1}{N-1}\sum^{N-1}_{i=1}\Bigg(\overline{\varepsilon}_{i-}^{\{N,\alpha\}}-\overline{\varepsilon}_{N-}^{\{N,\alpha\}}\Bigg),
\ee
and,
similarly,
\be\label{eq:deriv_N-centered_ener_DFT_right}
&&\dfrac{\partial
\mathscr{E}_+^{\{N,\alpha\}}}{\partial\alpha}
-\left.\dfrac{\partial E^{\{N,\alpha\}}_{\rm
Hxc+}[n]}{\partial \alpha}\right|_{n=
n_{\hat{\gamma}_{\rm s+}^{\{N,\alpha\}}}}
\nonumber\\
&=&
-\dfrac{1}{N+1}\sum^{N}_{i=1}\Bigg(\overline{\varepsilon}_{i+}^{\{N,\alpha\}}-\overline{\varepsilon}_{(N+1)+}^{\{N,\alpha\}}\Bigg).
\ee
By combining Eqs.~(\ref{eq:true_ener_from_left}),
(\ref{eq:true_ener_from_right}), and (\ref{eq:N-centered_ener_DFT_left})--(\ref{eq:deriv_N-centered_ener_DFT_right})
we can finally express 
the exact open-system energy in terms of the LZ-shifted KS orbital
energies as follows:  
\be\label{eq:opensys_energy_fromclosed_-}
E\left(\mathcal{N}\right)
&\overset{\ceil{\mathcal{N}}=N}{\equiv}&
 E\left(N-\alpha\right)
\nonumber\\
&
=&\sum^{N-1}_{i=1}\overline{\varepsilon}_{i-}^{\{N,\alpha\}}
+(1-\alpha)\overline{\varepsilon}_{N-}^{\{N,\alpha\}}
\nonumber\\
&&-\dfrac{\alpha(1-  \alpha)}{N}\left.\dfrac{\partial E^{\{N,\alpha\}}_{\rm
Hxc-}[n]}{\partial \alpha}\right|_{n=
n_{\hat{\gamma}_{\rm s-}^{\{N,\alpha\}}}},
\ee
and
\be\label{eq:opensys_energy_fromclosed_+}
E\left(\mathcal{N}\right)
&\overset{\floor{\mathcal{N}}=N}{\equiv}&
 E\left(N+\alpha\right)
\nonumber\\
&=&\sum^{N}_{i=1}\overline{\varepsilon}_{i+}^{\{N,\alpha\}}
+\alpha\overline{\varepsilon}_{(N+1)+}^{\{N,\alpha\}}
\nonumber\\
&&+
\dfrac{\alpha(1-  \alpha)}{N}
\left.\dfrac{\partial E^{\{N,\alpha\}}_{\rm
Hxc+}[n]}{\partial \alpha}\right|_{n=
n_{\hat{\gamma}_{\rm s+}^{\{N,\alpha\}}}}.
\ee
Eqs.~(\ref{eq:opensys_energy_fromclosed_-}) and
(\ref{eq:opensys_energy_fromclosed_+}) are the central result of this
work.\\ 

As readily seen from these equations, the exact physical energies
are recovered by summing up the LZ-shifted occupied KS orbital energies
not only in the $N$-electron system
(i.e. when $\alpha=0$), as expected from Ref.~\cite{levy2014ground}, but
also in the $(N\pm1)$-electron ones (i.e. when $\alpha=1$). This is a
non-trivial result since the left/right $N$-centered ensemble densities do
not reduce
exactly to the physical ones when $\alpha=1$ [see the prefactors
$1\pm(\alpha/N)$ in Eqs.~(\ref{eq:true_dens_from_left}) and (\ref{eq:true_dens_from_right})].   
In addition, keeping in mind that the physical energies $E(N\pm\alpha)$
vary linearly with $\alpha$, we expect the total LZ-shifted KS
energies
\be\label{eq:total_LZ-shifted_ener_l}
\sum^{N-1}_{i=1}\overline{\varepsilon}_{i-}^{\{N,\alpha\}}
+(1-\alpha)\overline{\varepsilon}_{N-}^{\{N,\alpha\}}
\ee
and
\be\label{eq:total_LZ-shifted_ener_r}
\sum^{N}_{i=1}\overline{\varepsilon}_{i+}^{\{N,\alpha\}}
+\alpha\overline{\varepsilon}_{(N+1)+}^{\{N,\alpha\}}
,
\ee
for $N-\alpha$ and $N+\alpha$ electrons, respectively, to vary (at
least)
quadratically with $\alpha$. Linearity should then be recovered when
adding the ensemble Hxc first-order-derivative corrections [third terms
on the right-hand side of Eqs.~(\ref{eq:opensys_energy_fromclosed_-}) and
(\ref{eq:opensys_energy_fromclosed_+})]. This is illustrated in
Fig.~\ref{fig:noxcDD} with the Hubbard dimer model [see
Sec.~\ref{sec:hubbard} for further details].        

\subsection{Reformulation of the IP/EA theorem}\label{subsec:IP-EA_th}

As shown in Sec.~\ref{subsec:L-R-Nc-ensembles}, the left/right $N$-centered ensemble
energies (and their first-order derivatives in $\alpha$) give directly
access not only to the
physical energy for {\it any} fractional
electron number $\mathcal{N}$ (i.e. any value of $\alpha$), but also to the IP and the EA [see Eqs.~(\ref{eq:IP_from_n-centered_ens}) and
(\ref{eq:EA_from_n-centered_ens})]. From the KS-DFT expressions in
Eqs.~(\ref{eq:N-centered_ener_DFT_left})--(\ref{eq:deriv_N-centered_ener_DFT_right})
we obtain the following reformulation
of Janak's theorem~\cite{janak1978proof}, which holds for {\it any} value of $\mathcal{N}$ or,
equivalently, for {\it any} value of $\alpha$ in the range $0\leq
\alpha\leq 1$:
\be\label{eq:new_IP_th}
&&\dfrac{dE\left(\mathcal{N}\right)}{d\mathcal{N}}
\overset{\mathcal{N}=N-\alpha}{=}
-I^N
\nonumber\\
&=&\overline{\varepsilon}_{N-}^{\{N,\alpha\}}+\dfrac{(1-\alpha-N)}{N}\left.\dfrac{\partial E^{\{N,\alpha\}}_{\rm
Hxc-}[n]}{\partial \alpha}\right|_{n=
n_{\hat{\gamma}_{\rm s-}^{\{N,\alpha\}}}},
\ee
and
\be\label{eq:new_EA_th}
&&\dfrac{dE\left(\mathcal{N}\right)}{d\mathcal{N}}
\overset{\mathcal{N}=N+\alpha}{=}
-A^N
\nonumber\\
&=&\overline{\varepsilon}_{(N+1)+}^{\{N,\alpha\}}+\dfrac{(1-\alpha+N)}{N}\left.\dfrac{\partial E^{\{N,\alpha\}}_{\rm
Hxc+}[n]}{\partial \alpha}\right|_{n=
n_{\hat{\gamma}_{\rm s+}^{\{N,\alpha\}}}}.
\ee
This is the second key result of the paper.\\

In the limits $\mathcal{N}\rightarrow N\pm\eta$ where
$\eta\rightarrow 0^+$, which consists in taking
the $\alpha=0$ limit in both left and right $N$-centered ensembles, the
density becomes simply the conventional $N$-electron ground-state 
density: 
\be
n_{\hat{\gamma}_{\rm
s\pm}^{\{N,\alpha=0\}}}(\br)=n_{\hat{\Gamma}^N}(\br),
\ee
and the left/right ensemble Hxc functionals reduce to the
conventional ($N$-electron) ground-state one [see
Eq.~(\ref{eq:alpha_zero_limit_Hxcfun})].  
The KS potential in Eq.~(\ref{eq:sc-KS-L-R_eqs}) is unique {\it up to a
constant}, as we systematically search for the $N$- and $(N\pm1)$-electron ground states of a
non-interacting system in order to construct left or right
$N$-centered KS ensembles. As a result, $v^{\{N,\alpha=0\}}_{\rm s-}(\br)$
and $v^{\{N,\alpha=0\}}_{\rm s+}(\br)$ may differ by a constant.
However, once the LZ shift is applied, the potential becomes truly
unique [see Eq.~(\ref{eq:LZ_shift_L-R})], thus leading to 
\be
\overline{v}^{\{N,\alpha=0\}}_{\rm
s-}(\br)=\overline{v}^{\{N,\alpha=0\}}_{\rm s+}(\br).
\ee
Consequently, we have
\be
\overline{\varepsilon}_{i-}^{\{N,\alpha=0\}}=\overline{\varepsilon}_{i+}^{\{N,\alpha=0\}},
\ee
and, in particular, 
\be
\overline{\varepsilon}_{(N+1)+}^{\{N,\alpha=0\}}-
\overline{\varepsilon}_{N-}^{\{N,\alpha=0\}}
&=&\overline{\varepsilon}_{(N+1)\pm}^{\{N,\alpha=0\}}-             \overline{\varepsilon}_{N\pm}^{\{N,\alpha=0\}}
\nonumber\\
&=&{\varepsilon}_{N+1}-{\varepsilon}_{N}
,
\ee
where ${\varepsilon}_{N}$ and ${\varepsilon}_{N+1}$ are,
respectively, the energies
of the HOMO and LUMO obtained from a regular $N$-electron KS-DFT
calculation. Note that the LZ shift does not affect the HOMO-LUMO gap. Therefore, according to the IP/EA expressions in
Eqs.~(\ref{eq:new_IP_th}) and (\ref{eq:new_EA_th}), the exact fundamental gap can
be written as follows:
\be
I^N-A^N&=&
{\varepsilon}_{N+1}-{\varepsilon}_{N}
+\dfrac{(N+1)}{N}
\left.\dfrac{\partial E^{\{N,\alpha\}}_{\rm
Hxc+}[n_{\hat{\Gamma}^N}]}{\partial \alpha}\right|_{\alpha=0}
\nonumber\\
&&+\dfrac{(N-1)}{N}\left.\dfrac{\partial E^{\{N,\alpha\}}_{\rm
Hxc-}[n_{\hat{\Gamma}^N}]}{\partial \alpha}
\right|_{\alpha=0},
\ee
where the last two terms on the right-hand side play all together the role of the
derivative discontinuity correction to the gap in conventional KS-DFT.
Note that, according to Eqs.~(\ref{eq:left_from_two_weight}) and
(\ref{eq:right_from_two_weight}), the left and right $N$-centered
ensemble Hxc functionals are connected to 
the original $N$-centered
ensemble functional 
$E^{\left\{N,\xi_-,\xi_+\right\}}_{\rm Hxc}[n]$ of Ref.~\cite{senjean2018unified} (the latter
describes the Hxc energy of the ensemble defined in Eq.~(\ref{eq:original_N-centered_ens_twoweights}))  
as follows:
\be\label{eq:Hxc_left_from_twoweight}
E^{\{N,\alpha\}}_{\rm
Hxc-}[n]=E_{\rm
Hxc}^{\left\{N,\xi_-=\frac{N\alpha}{N-1},\xi_+=0\right\}}[n],
\ee
and
\be\label{eq:Hxc_right_from_twoweight}
E^{\{N,\alpha\}}_{\rm
Hxc+}[n]=E_{\rm
Hxc}^{\left\{N,\xi_-=0,\xi_+=\frac{N\alpha}{N+1}\right\}}[n]
.
\ee
Thus we recover the compact expression~\cite{senjean2018unified}:   
\be
I^N-A^N&=&
{\varepsilon}_{N+1}-{\varepsilon}_{N}
+\left.\dfrac{\partial E^{\{N,\xi,\xi\}}_{\rm
Hxc}[n_{\hat{\Gamma}^N}]}{\partial \xi}\right|_{\xi=0}.
\ee
Interestingly, designing density-functional approximations for the
original (two-weight) $N$-centered
ensembles benefits automatically to the left/right variants of the theory
through Eqs.~(\ref{eq:Hxc_left_from_twoweight}) and (\ref{eq:Hxc_right_from_twoweight}).  
 
\subsection{Left or right ?}\label{subsec:left_or_right}

For clarity and convenience, we have derived explicitly {\it both} left and right variants of
$N$-centered ensemble DFT. This is in principle not necessary as we may
opt for one of the ensemble (left or right) and simply increment or
decrement the central number of electrons.    
As illustrated in Fig.~\ref{fig:l-r-schematics}, an open $\mathcal{N}$-electron system
with $\ceil{\mathcal{N}}=N$ can
be described either by a left ensemble centered on $N$ {\it or}
a right ensemble centered on $(N-1)$. In other words, one may express
the number of electrons as
\be
\mathcal{N}=N-\alpha,
\ee
or, equivalently,
\be
\mathcal{N}=N-1+\alpha',
\ee
where $\alpha'=1-\alpha$.
The two descriptions should of
course lead to the same physical density and energy. It is actually
quite simple to obtain 
the left ensemble Hxc density functional
from the right one, and {\it vice
versa}. Indeed, according to Eqs.~(\ref{eq:link_left_right_ensembles})
and (\ref{eq:LL_L-R_fun}),
\be
&&F_-^{\{N,1-\alpha\}}[n]=\dfrac{N}{N-1}
\nonumber\\
&&\times
\min_{\frac{N}{N-1}\hat{\gamma}_+^{\{N-1,\alpha\}}\rightarrow n}\left\{
\tra{\hat{\gamma}_+^{\{N-1,\alpha\}}\left(\hat{T}+\hat{W}_{\rm          ee}\right)}
\right\}
\nonumber\\
&&
=\dfrac{N}{N-1}
\min_{\hat{\gamma}_+^{\{N-1,\alpha\}}\rightarrow
\frac{(N-1)n}{N}}\left\{
\tra{\hat{\gamma}_+^{\{N-1,\alpha\}}\left(\hat{T}+\hat{W}_{\rm          ee}\right)}
\right\}
\nonumber\\
&&
=\dfrac{N}{N-1}F_+^{\{N-1,\alpha\}}\left[\dfrac{(N-1)n}{N}\right],
\ee
where $n$ is an $N$-electron density. Similarly, we can show that, if $n$ is an
$(N-1)$-electron density, then   
\be
F_+^{\{N-1,\alpha\}}\left[n\right]=\dfrac{N-1}{N}F_-^{\{N,1-\alpha\}}\left[\dfrac{Nn}{(N-1)}\right].
\ee
As these scaling relations apply also to the non-interacting kinetic energy, we
conclude that
\be\label{eq:left_from_right_Hxc_fun}
E^{\{N,1-\alpha\}}_{\rm Hxc-}\left[n\right]=
\dfrac{N}{N-1}
E^{\{N-1,\alpha\}}_{\rm Hxc+}\left[\dfrac{(N-1)n}{N}\right],
\ee
and
\be\label{eq:right_from_left_Hxc_fun}
E_{\rm Hxc+}^{\{N-1,\alpha\}}\left[n\right]=\dfrac{N-1}{N}E_{\rm Hxc-}^{\{N,1-\alpha\}}\left[\dfrac{Nn}{(N-1)}\right].
\ee
In summary, a single (left or right) $N$-centered Hxc functional is needed for
describing the variation of the electron number between two integers.
One functional is obtained from the other {\it via} the scaling relations
of Eqs.~(\ref{eq:left_from_right_Hxc_fun}) and
(\ref{eq:right_from_left_Hxc_fun}).\\

Note that these relations enable to understand why, in the
$\alpha=1$ limit of left and
right $N$-centered ensembles, the sum of the LZ-shifted occupied KS orbital energies matches
the physical $(N-1)$- and $(N+1)$-electron energies, respectively [see
Eqs.~(\ref{eq:Hxc_pot_d_over_dn}), (\ref{eq:LZ_shift_L-R}), 
(\ref{eq:opensys_energy_fromclosed_-}), and (\ref{eq:opensys_energy_fromclosed_+})].

\subsection{Density-driven correlations in $N$-centered
ensembles}\label{subsec:DDeffects}

Very recently, Gould and Pittalis~\cite{gould2019density,gould2020density} revealed that direct approximations to
Gross--Oliveira--Kohn (GOK) ensemble DFT, where the ensemble
consists of $N$-electron ground and excited states~\cite{gross1988density}, miss an important
correlation effect that they refer to as {\it density-driven} correlation. 
The latter originates from the
deviation in density of the individual KS states within the
ensemble from the true
interacting ones, 
as shown explicitly by one of the author~\cite{fromager2020individual}. 
One may naturally wonder if this kind of correlation
exists also in $N$-centered ensemble DFT. In the following, we will
briefly explain how density-driven correlations can be defined in this
context. We leave their detailed study for future work.\\

Let us follow the derivation of Ref.~\cite{fromager2020individual} and  
adapt it to (say left) $N$-centered
ensembles. We start by extracting the individual $M$-electron ($M=N$ or
$N-1$) total energies $E(M)=\tra{\hat{\Gamma}^M\hat{H}}$. For that
purpose we use 
Eqs.~(\ref{eq:GammaN_from_L-R_ens}) and (\ref{eq:GammaNpm1_from_LorR_ens}),
and the {\it variational} $N$-centered ensemble energy expression of
Eq.~(\ref{eq:KS1}), thus leading to the following (exact) expression:   
\be\label{eq:ind_ener_from_L}
E(M)&=&
\tra{\hat{\gamma}^{\{N,\alpha\}}_{{\rm
s}-,M}\hat{h}
}
+E_{{\rm Hxc-},M}^{\{N,\alpha\}}\left[n_{\hat{\gamma}_{\rm
s-}^{\{N,\alpha\}}},n_{\hat{\gamma}^{\{N,\alpha\}}_{{\rm
s}-,M}}\right],
\nonumber\\
\ee
where $\hat{\gamma}^{\{N,\alpha\}}_{{\rm
s}-,M}$ is the $M$-electron component of the KS 
$N$-centered ensemble density matrix operator:
\be
\hat{\gamma}_{\rm s-}^{\{N,\alpha\}}=(1-\alpha)\hat{\gamma}^{\{N,\alpha\}}_{{\rm
s}-,N}+\dfrac{N\alpha}{N - 1}\hat{\gamma}^{\{N,\alpha\}}_{{\rm
s}-,N-1}.
\ee
The individual $N$- and $(N-1)$-electron Hxc bifunctionals in
Eq.~(\ref{eq:ind_ener_from_L}) read
\be
&&E_{{\rm
Hxc-},N}^{\{N,\alpha\}}\left[n,n^N\right]=
\left(
1-
\alpha\dfrac{\partial
}{\partial\alpha}
\right)
E_{{\rm
Hxc}-}^{\{N,\alpha\}}[n]
\nonumber\\
&&+\int\dr\dfrac{\delta  E_{{\rm
Hxc}-}^{\{N,\alpha\}}[n]}{\delta n(\br)}\left(
n^N
(\br)-n(\br)\right), 
\ee  
and
\be
&&E_{{\rm
Hxc-},N-1}^{\{N,\alpha\}}\left[n,n^{N-1}\right]=
\nonumber\\
&&\dfrac{N-1}{N}\left(
1+(1-
\alpha)\dfrac{\partial 
}{\partial\alpha}
\right)E_{{\rm
Hxc}-}^{\{N,\alpha\}}[n]
\nonumber\\
&&+\int\dr\dfrac{\delta  E_{{\rm
Hxc}-}^{\{N,\alpha\}}[n]}{\delta n(\br)}\left(
n^{N-1}
(\br)-\dfrac{(N-1)}{N}n(\br)\right), 
\nonumber\\
\ee
respectively.
One should then realize that the density constraint in Eq.~(\ref{eq:Ts_L-R})
does {\it not} necessarily imply that the individual KS densities match the true 
interacting ones:
\be
n_{\hat{\gamma}^{\{N,\alpha\}}_{{\rm
s}-,M}}\neq n_{\hat{\Gamma}^M}.
\ee
This can be illustrated easily with the
asymmetric Hubbard dimer model, which will be studied in detail 
in Sec.~\ref{sec:hubbard}. In this model, the one-electron ground-state density (which is an
atomic site occupation) reads  
\be\label{eq:one-elec_dens_dimer}
n^{M=1}=\dfrac{1}{2}+\dfrac{\Delta v}{2\sqrt{4t^2+\Delta v^2}},
\ee
where $t$ is the strength of the kinetic energy and $\Delta v$ is the analog of the local potential (here it is
just a number that controls the asymmetry in the dimer). When the dimer
is asymmetric (i.e. $\Delta v\neq 0$), the Hxc potential [which is the
difference in potential between  
the KS noninteracting $N$-centered ensemble and the interacting one] is
nonzero. In the weakly correlated regime, this can be readily seen from 
Eq.~(\ref{eq:Hx-}).  
Therefore, according to Eq.~(\ref{eq:one-elec_dens_dimer}), the true and KS one-electron states 
[each of them being a component of a (left) $N$-centered ensemble with
$N=2$] do not have the same
density. If we now return to the general case, it is then relevant to decompose
each individual
correlation energy
into state-driven (SD) and density-driven (DD) contributions, by analogy
with GOK ensembles~\cite{fromager2020individual}:
\be
E_{{\rm c-},M}^{\{N,\alpha\}}\left[n_{\hat{\gamma}_{\rm
s-}^{\{N,\alpha\}}},n_{\hat{\gamma}^{\{N,\alpha\}}_{{\rm
s}-,M}}\right]\equiv E_{{\rm c-},M}^{\{N,\alpha\}{\rm SD}}+E_{{\rm
c-},M}^{\{N,\alpha\}\rm DD},
\ee
where 
\be
E_{{\rm c-},M}^{\{N,\alpha\}{\rm SD}}:=E_{{\rm c-},M}^{\{N,\alpha\}}\left[n_{\hat{\gamma}_{\rm
s-}^{\{N,\alpha\}}},n_{\hat{\Gamma}^M}
\right].
\ee
While we do not expect the conventional ground-state functional $E_{\rm
Hxc}[n]$ to be a good approximation to (neutrally) excited-state Hxc energies, one may wonder how good the approximation $E_{{\rm
Hxc-},M}^{\{N,\alpha\}}\left[n,n^M\right]\approx E_{\rm
Hxc}\left[n^M\right]$ is since an
$N$-centered ensemble consists of ground states only. Answering this
question is expected to pave the way toward the rationalization and development
of density-functional approximations that incorporate derivative
discontinuity corrections. Work is currently in
progress in this direction.   

\subsection{Comparison with conventional DFT for open
systems}\label{subsec:comparison_with_Levy-Perdew-approach}

In standard PPLB-DFT for open systems~\cite{perdew1982density}, the
density $n$ of the true physical open system is used as basic
variable. It gives, 
by
integration,
the total (fractional) number $\mathcal{N}$ of electrons: 
\be
\mathcal{N}=\int\dr\,n(\br).
\ee
The Hxc energy is then obtained from the universal ground-state
functional $E_{\rm Hxc}[n]$ which is defined over the domain of
densities that can integrate to any {\it integral or fractional} number of
electrons. 
Despite its formal beauty, this grand canonical formulation
of DFT is not necessarily the most appealing one when it comes to
perform practical calculations. Indeed, modeling $E_{\rm
Hxc}[n]$ accurately for any number of electrons (including fractional ones) is
not straightforward. Most importantly, the model should be able to
reproduce the
derivative discontinuity that the functional is expected to exhibit when crossing an
integer~\cite{perdew1983physical}. Note that, for open systems, the KS
potential is truly unique (not up to a constant anymore). Indeed,
the chemical potential has to be adjusted such that the grand canonical
ground-state energies of
the $(N-1)$- and $N$-electron systems [we assume that
$N-1<\mathcal{N}<N$] are {\it equal}.\\  

The situation is completely different in (say) left $N$-centered
ensemble DFT, where we use two variables instead. The first one is
the left $N$-centered
ensemble density that integrates to the
so-called central number
$N=\ceil{\mathcal{N}}$ of electrons. 
The second variable $\alpha=N-\mathcal{N}$
is the deviation of the central number from the true electron number $\mathcal{N}$.
Even though alternative ensemble DFT
approaches~\cite{kraisler2013piecewise,gorling2015exchange} do not rely
on a 
centered density, they use information about the $(N-1)$ and $N$-electron
systems (i.e. systems with an {\it integral} number of electrons) as $N$-centered ensemble DFT does. They are similar from this
point of view.\\
 
Since a density $n$ that integrates to $N$ can be reproduced by either a pure
$N$-electron ground state or a left $N$-centered ensemble, it is
essential to construct a functional $E_{\rm
Hxc-}^{\{N,\alpha\}}\left[n\right]$ that is both density- {\it and}
$\alpha$-dependent. At first sight, this version of DFT for open
systems looks much more
complicated than the PPLB one. What makes the $N$-centered
formulation appealing is that the infamous derivative discontinuity 
is now obtained through the derivative in $\alpha$ of the $N$-centered
ensemble density-functional Hxc energy. As a result, if one can
model the dependence in $\alpha$ of the latter functional, one can in
principle
solve the derivative discontinuity problem.

\section{Application to the Hubbard dimer}\label{sec:hubbard}

\subsection{Hamiltonian and density functionals}
A proof of concept study of the (not necessarily symmetric) Hubbard
dimer is presented in this section. 
This is one of the simplest solvable models 
exhibiting a nontrivial interplay 
between electron-electron
interaction and inhomogeneity. It is often used as a lab for testing new
ideas in DFT, gaining more insight into those and their approximate
formulations~\cite{fuks2013time,
carrascal2015hubbard,
deur2017exact,
senjean2017local,
deur2018exploring,
carrascal2018linear,
vanzini2018spectroscopy,
senjean2018unified,
li2018density,
deur2019ground}.
In this model,
the {\it ab initio} Hamiltonian $\hat{H}$ is simplified as follows:
\be\label{eq:Hamil_Hubbard_dimer_model}
\hat{T} &\rightarrow& -t
\sum_{\sigma=\uparrow\downarrow}(\hat{c}^\dagger_{0\sigma}\hat{c}_{1\sigma} + 
\hat{c}^\dagger_{1\sigma}\hat{c}_{0\sigma}),\hspace{0.2cm} \hat{W}_{\rm ee}\rightarrow
U\sum^1_{i=0}\hat{n}_{i\uparrow}\hat{n}_{i\downarrow}, 
\nonumber\\
\hat{V}_{\rm ext}&\rightarrow&\Delta v_{\rm ext}(\hat{n}_1 -
\hat{n}_0)/2,\hspace{0.3cm}\hat{n}_{i\sigma}=\hat{c}^\dagger_{i\sigma}\hat{c}_{i\sigma},
\ee
where $\hat{n}_i=\sum_{\sigma=\uparrow\downarrow}\hat{n}_{i\sigma}$ is
the density operator on site $i$ ($i=0,1$). Note that the external potential reduces to a single number $\Delta v_{\rm
ext}$ which controls the asymmetry of the dimer. 
The density also reduces 
to a single number $n=n_0$ which is the occupation of site 0 given that
$n_1 = N-n$ for an $N$-centered ensemble. In the following, the central
number of electrons will be {\it fixed to $N=2$}
and the hopping parameter will be set to $t = 1$.
All density functionals can be derived analytically except the
correlation ones that can be computed to arbitrary accuracy {\it via}
Lieb
maximizations~\cite{deur2017exact,senjean2018unified,deur2019ground}.\\

As shown in Appendix~\ref{app:derivations}, within left/right
$N$-centered ensemble DFT, the exact ground-state energy of an open system
with $\mathcal{N}_{\pm} = 2\pm \alpha$ electrons reads
\be\label{eq:opensys_energy_Hubbard}
&&E\left(\mathcal{N}_\pm\right)\equiv
\Bigg[
 \dfrac{2\pm\alpha}{2}\Big(
T^{\{N=2,\alpha\}}_{\rm s\pm}(n)+E^{\{N=2,\alpha\}}_{\rm Hxc\pm}(n)
\nonumber\\
&&+\Delta v_{\rm
ext}\times\left(1-n\right)
\Big) \pm \dfrac{\alpha(1-\alpha)}{2}
\dfrac{\partial T^{\{N=2,\alpha\}}_{\rm s\pm}(n)}{\partial\alpha} \nonumber\\
&&
\pm \dfrac{\alpha(1-\alpha)}{2}
\dfrac{\partial E^{\{N=2,\alpha\}}_{\rm Hxc\pm}(n)}{\partial\alpha}
\Bigg]_{n=n^{\{N=2,\alpha\}}_\pm\left(\Delta v_{\rm
ext}\right)}.
\ee
According
to Eqs.~(\ref{eq:left_from_two_weight}),
(\ref{eq:right_from_two_weight}), (\ref{eq:Ts_L-R}), and (\ref{eq:EEXX_l-r}), 
the expressions for the 
left/right $N$-centered
ensemble non-interacting kinetic and Hx energy 
functionals can
be obtained from the two-weight $N$-centered ensemble
functionals of our previous work~\cite{senjean2018unified}.
Indeed, the left
functionals correspond to $\xi_-=\frac{N\alpha}{N-1}$ and
$\xi_+=0$, thus leading to
\be
T^{\{N=2,\alpha\}}_{\rm s-}(n)&=&
- 2t \sqrt{1 - (n - 1)^2}=T^{N=2}_{\rm s}(n),\label{eq:Ts-}\\
\Delta v_{\rm KS-}^{\{N=2 , \alpha\}}(n)&=& \dfrac{2t(n-1)}{\sqrt{1 - (1-n)^2}},\label{eq:vKS-}
\ee 
where $\Delta v_{\rm KS-}^{\{N=2 , \alpha\}}(n)=\partial
T^{\{N=2,\alpha\}}_{\rm s-}(n)/\partial n$ is the KS potential
for the left ensemble, 
and
\be\label{eq:Hx-}
E^{\{N=2,\alpha\}}_{\rm Hx-}(n)=\dfrac{U(1-\alpha)}{2}
\left(1+\left({n-1}\right)^2\right).
\ee
Similarly for the right functionals, we have $\xi_-=0$ and
$\xi_+=\frac{N\alpha}{N+1}$, thus leading to
\be
T^{\{N=2,\alpha\}}_{\rm s+}(n)&=&- 2t \sqrt{\left(\dfrac{2\alpha}{3} - 1\right)^2 - (n - 1)^2},\label{eq:Ts+}\\
\Delta v_{\rm KS+}^{\{N=2 , \alpha\}}(n)&=& \dfrac{2t(n-1)}{\sqrt{\left(\dfrac{2\alpha}{3} - 1\right)^2 - (n - 1)^2}},\label{eq:vKS+}
\ee
and
\be\label{eq:Hx+}
E^{\{N=2,\alpha\}}_{\rm Hx+}(n)=\dfrac{U}{2}\left[1 +\dfrac{\alpha}{3}
+\dfrac{9\left(1 -\alpha\right)}{\left(2\alpha - 3\right)^2} \left(
{n-1}\right)^2\right].
\ee

As readily seen from Eq.~(\ref{eq:vKS+}), the non-interacting
$v$-representability condition for the right $N$-centered ensemble is
$\alpha$-dependent:
\be
\dfrac{2\alpha}{3} < n < 2 - \dfrac{2\alpha}{3}.
\ee
Various density-functional approximations are tested in the
following. For simplicity, we will restrict our study to
functional-driven errors~\cite{kim2013understanding} which means that all density-functional
energies will be computed with the {\it exact} $N$-centered ensemble
densities. The investigation of density-driven errors~\cite{kim2013understanding} is left for future
work. Let us start with two approximations that would be applicable also to 
{\it ab initio} problems. The simplest one consists in neglecting
correlation effects and using a regular ($\alpha$-independent) Ground-State exchange
functional. In this model, we simply have to consider the $\alpha=0$ limit of
Eqs.~(\ref{eq:Hx-}) and (\ref{eq:Hx+}). The approximation will be
referred to as GS-Hx. A conventional ($\alpha$-independent) Ground-State
correlation functional might then be added, thus leading 
to a second
approximation referred to as GS-Hxc. In this context, we use the correlation functional
of Carrascal {\it et al.}~\cite{carrascal2015hubbard}
which has been parameterized on the two-electron Hubbard dimer. 
Finally, in order to investigate the importance of the
$\alpha$-dependence in both exchange and correlation ensemble energies,
we consider two additional approximations. In the first one, which is referred to
as {\it ensemble exact exchange} (EEXX), Eqs.~(\ref{eq:Hx-}) and
(\ref{eq:Hx+}) are employed, and correlation effects are neglected. Adding the
Ground-State correlation functional of Carrascal {\it et
al.}~\cite{carrascal2015hubbard} leads to our last (so-called GS-c)
approximation. The four approximations are summarized in
Tab.~\ref{tab:approx}.
\begin{table}
\caption{Density-functional approximations tested in this
work.}
\label{tab:approx}
\begin{tabular}{l|l|l}
\hline \\
Acronym  ~~~~ & $E^{\{N=2,\alpha\}}_{\rm Hxc\pm}(n)\approx$ & References \\ [0.2cm]
\hline \\
GS-Hx & $E_{\rm
Hx}(n)$ &  Eq.~(\ref{eq:Hx-}) with $\alpha=0$\\ [0.2cm]
GS-Hxc & $E_{\rm
Hx}(n)+E_{\rm c}(n)$ &  Ref.~\cite{carrascal2015hubbard} \\ [0.2cm]
EEXX & $E^{\{N=2,\alpha\}}_{\rm Hx\pm}(n)$ & Eqs.~(\ref{eq:Hx-}), (\ref{eq:Hx+}) \\ [0.2cm]
GS-c & $E^{\{N=2,\alpha\}}_{\rm Hx\pm}(n)+E_{\rm c}(n)$ &
Eqs.~(\ref{eq:Hx-}), (\ref{eq:Hx+}), Ref.~\cite{carrascal2015hubbard}\\ [0.2cm]
\hline
\end{tabular}
\end{table}

\subsection{Influence of the ensemble derivative-in-$\alpha$ Hxc density
functional
}

\begin{figure}
\centering
\resizebox{\columnwidth}{!}{
\includegraphics[scale=1]{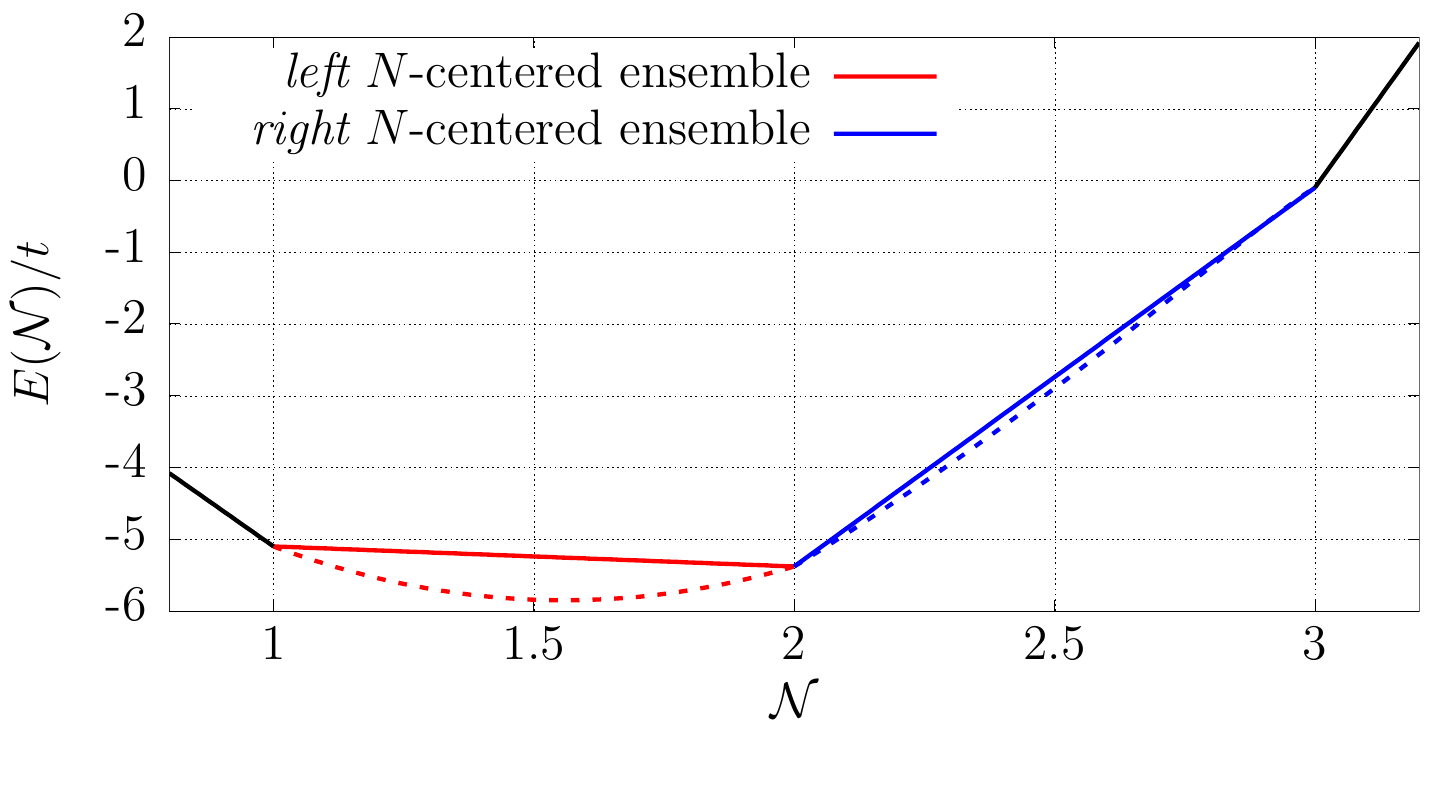}
}
\caption{Total LZ-shifted KS (dashed lines) and 
exact (full lines) energies plotted as functions of the electron number for the asymmetric Hubbard dimer with
$\Delta v_{\rm ext}/t = 10$ and $U/t = 5$. Results are shown for the value $N=2$
of the central number. See Eqs.~(\ref{eq:total_LZ-shifted_ener_l}) and
(\ref{eq:total_LZ-shifted_ener_r}), and the text that follows for further details.}
\label{fig:noxcDD}
\end{figure}

As shown in Eqs.~(\ref{eq:opensys_energy_fromclosed_-})
and (\ref{eq:opensys_energy_fromclosed_+}),
the true physical
energy of an open system
cannot be expressed solely
as a function of the shifted KS orbital energies.
An additional
ensemble Hxc first-order-derivative correction
has to be accounted for as soon as the true number
of electrons deviates from an integer.
In Fig.~\ref{fig:noxcDD},
we plot the exact expressions
for the true physical energy
using the left and right ($N$=2)-centered ensembles,
with and without the ensemble derivative-in-$\alpha$ Hxc corrections.
As expected, the exact energy is a piecewise linear function of the electron number.
Removing the derivative-in-$\alpha$ correction 
induces curvature. Hence, in the language of left/right $N$-centered ensemble DFT,
describing the piecewise linearity of the energy consists in modeling
the dependence in $\alpha$ of the ensemble Hxc density functional.
As clearly shown in Fig.~\ref{fig:noxcDD},
no derivative-in-$\alpha$ correction is needed when the number of electrons
is an integer. This is a direct consequence of the scaling relations in
Eqs.~(\ref{eq:left_from_right_Hxc_fun}) and (\ref{eq:right_from_left_Hxc_fun}), and the fact that, by construction, summing up the
LZ-shifted occupied KS orbital energies
gives the exact
energy for an integral number of electrons~\cite{levy2014ground}.

\subsection{Symmetric case}

\begin{figure}
\centering
\resizebox{!}{0.65\textwidth}{
\includegraphics[scale=1]{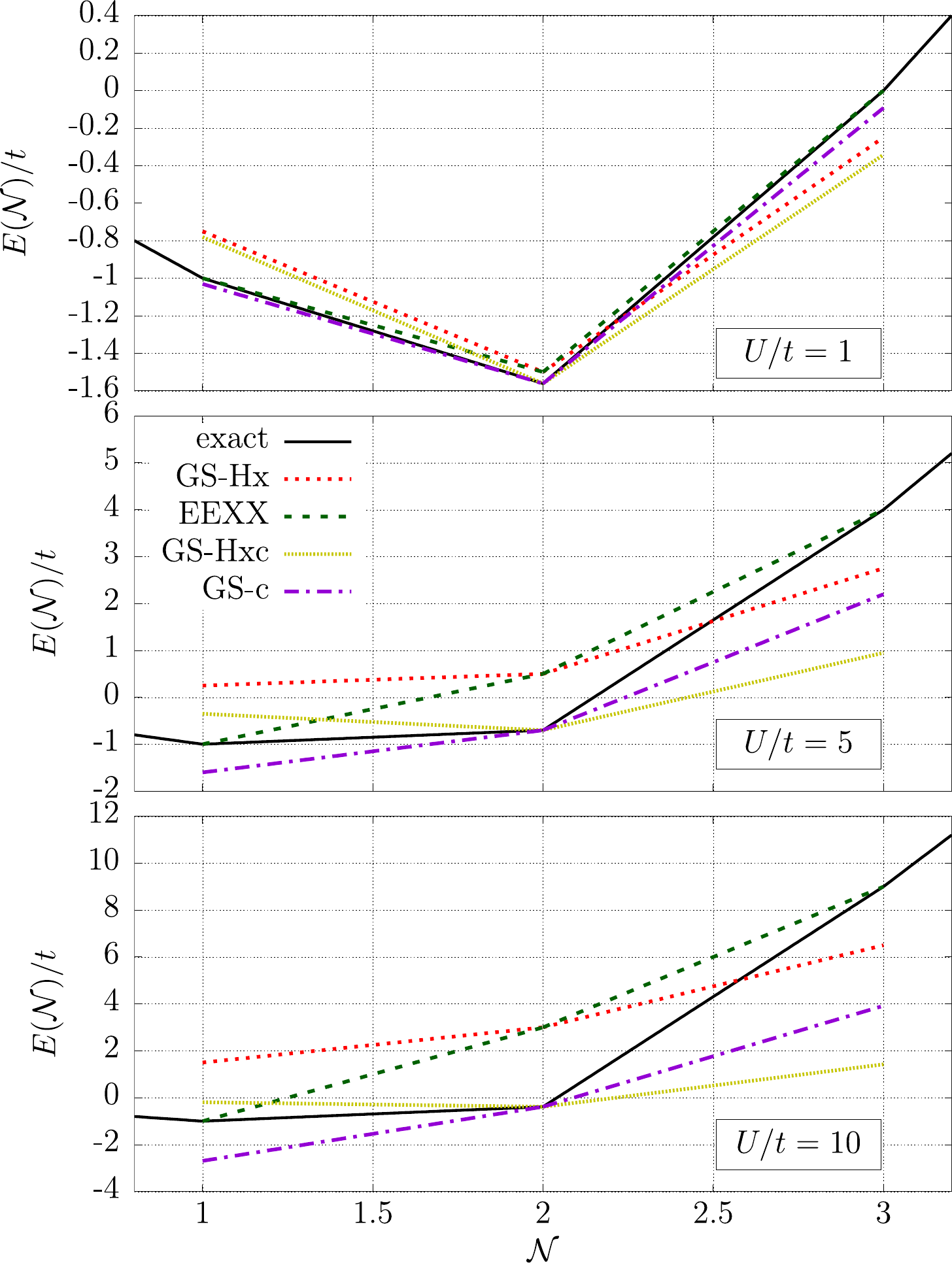}
}
\caption{Exact and approximate energies of the symmetric ($\Delta v_{\rm
ext}/t = 0$) Hubbard dimer plotted as functions of the electron number
$\mathcal{N}$ (or, equivalently, as functions of $\alpha$). The central
number of electrons is $N=2$. Results are
shown for $U/t = 1$, 5 and 10. Approximations are detailed in
Tab.~\ref{tab:approx}.}
\label{fig:symmetric}
\end{figure}

In this section, we investigate the performance of the various
approximations (see Tab.~\ref{tab:approx}) in the symmetric
dimer
($\Delta v_{\rm ext}/t = 0$)
for different values of $U/t$, as shown in Fig.~\ref{fig:symmetric}.
As expected, the GS-Hxc energy is always
below the GS-Hx one, because the correlation energy functional
is always negative (and there is no additional derivative-in-$\alpha$ correction). This statement also holds
when comparing GS-c with EEXX.
Since we use 
the highly accurate 
parametrization of Carrascal {\it et al.}~\cite{carrascal2015hubbard}
for the two-electron GS correlation energy functional,
the GS-Hxc and GS-c approximations are on top of the exact curve
for $\mathcal{N} = 2$.
This is definitely not the case for the 
GS-Hx approximation
which describes neither the correlation effects nor the 
$\alpha$-dependence. This is the reason why GS-Hx performs so poorly for
any $\mathcal{N}$ value, especially when $U/t$ is large. Note that
it also gives a wrong energy for $\mathcal{N}=1$. Indeed, 
even if there is only one electron (and therefore no correlation), the Hx part
of the functional is still $\alpha$-independent within GS-Hx, and
therefore only meaningful when $\mathcal{N} = 2$ (i.e. $\alpha=0$).\\

Returning to the
GS-Hxc and GS-c approximations, they are 
both exact for $\mathcal{N} = 2$.
When deviating from this central electron
number, the $\alpha$-independent
correlation functional
of Carrascal {\it et al.} becomes an approximation, as one
cannot expect the correlation energy to be
the same for systems with different numbers of electrons.
In contrast, EEXX
is exact for $\mathcal{N}=1$, by construction. 
It is also exact for $\mathcal{N} = 3$, which is 
due to the particle-hole symmetry of the model.
In the range $1 < \mathcal{N} < 3$, EEXX is not exact anymore
due to the lack of the $\alpha$-dependent correlation functional.
The effect of the latter on the true physical energy 
can be directly evaluated by
comparing EEXX to the exact curve.\\

Interestingly, all the approximations
give physical energies that vary linearly with $\mathcal{N}$ (or
$\alpha$). As shown in Appendix~\ref{app:symmetric}, this is due to the
fact that, in the symmetric dimer, the left and right ($N=2$)-centered
ensemble densities are equal to $n=1$. 

\subsection{Asymmetric case}

Let us now investigate the asymmetric case (we choose $\Delta v_{\rm
ext}/t = 5$) for different values of $U/t$.
\begin{figure}
\centering
\resizebox{\columnwidth}{!}{
\includegraphics[scale=1]{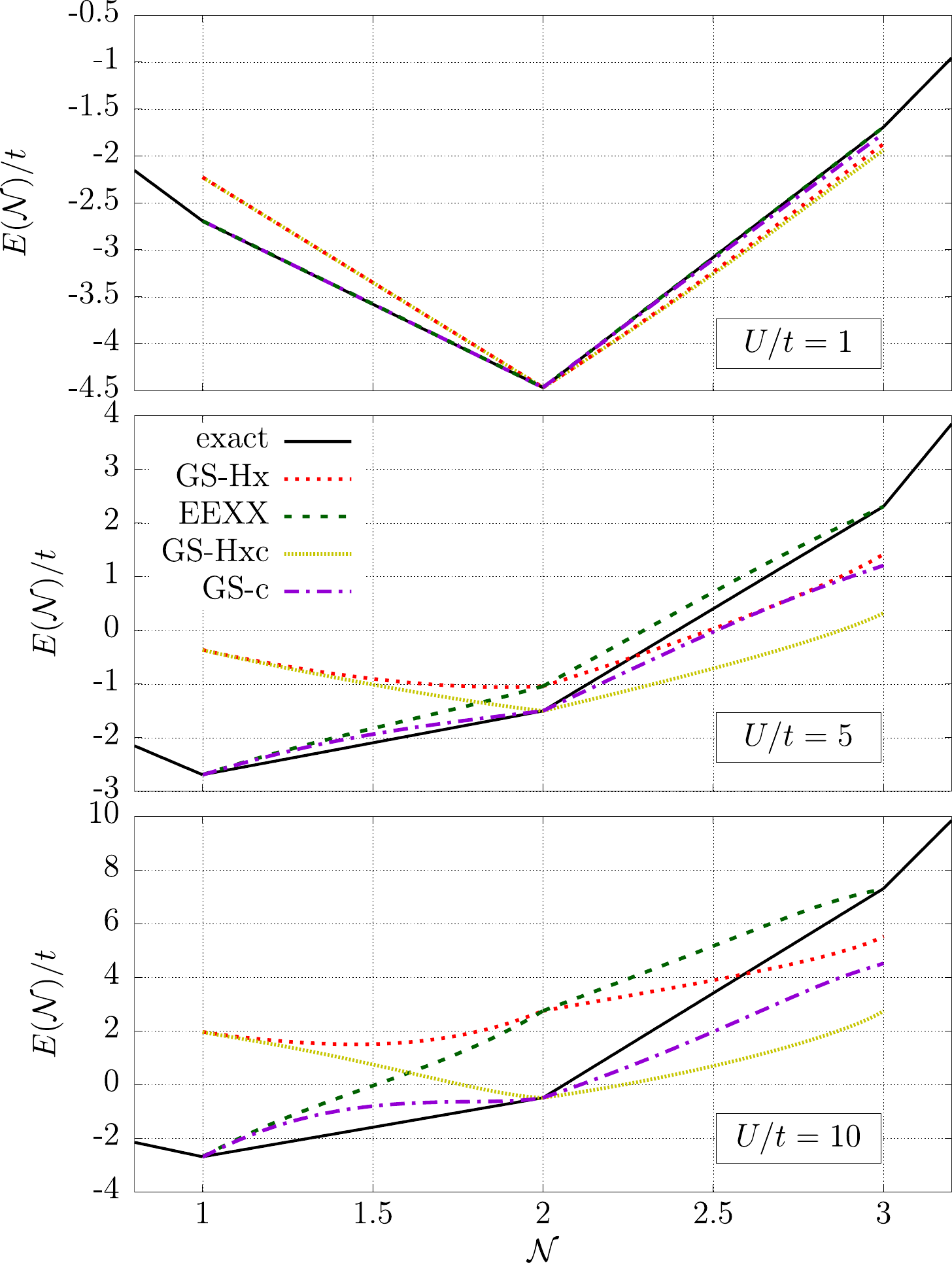}
}
\caption{
Exact and approximate energies of the asymmetric ($\Delta v_{\rm ext}/t = 5$) Hubbard dimer plotted as functions of the electron number
$\mathcal{N}$ (or, equivalently, as functions of $\alpha$). The central
number of electrons is $N=2$. Results are
shown for $U/t = 1$, 5 and 10. Approximations are detailed in
Tab.~\ref{tab:approx}.
}
\label{fig:asymmetric}
\end{figure}
As shown in Fig.~\ref{fig:asymmetric},
unlike in the symmetric case, approximate energies exhibit
curvature in $\mathcal{N}$.
In the particular case of GS-Hx and GS-Hxc (which are
$\alpha$-independent functionals), the curvature
is due to the $\alpha$-dependence in
both the non-interacting kinetic energy density functional and the
interacting $N$-centered ensemble density.
In the weakly-correlated regime ($U/t=1$ and $U/\Delta v_{\rm ext} = 1/5$)
one cannot distinguish the exact curve from the EEXX one,
showing that the
ensemble ($\alpha$-dependent) correlation density functional
is equal to 0 in this regime, for any $\alpha$.
This is due to the fact that the ensemble densities
reach the border of their non-interacting
$v$-representability
domain when $|\Delta v_{\rm ext}| \gg U$ and $|\Delta v_{\rm ext}| \gg
t$. For this particular value of the ensemble density,
the ensemble correlation density functional vanishes.
Note that the same behaviour
occurred in the original $N$-centered ensemble DFT,
for which the explanation can be found
in Appendix~C.3 of our previous work~\cite{senjean2018unified}.
Because the $v$-representability domain of
the left $N$-centered ensemble density is $\alpha$-independent,
the two-electron ($\alpha = 0$) correlation density functional
is also very close to 0 in between
$1 \leq \mathcal{N} \leq 2$, such that GS-c is now
on top of the exact
curve just like EEXX. 
It also explains why GS-Hxc and GS-Hx give the same
true physical energy here but, in contrast
to EEXX and GS-c, they are not exact thus highlighting the importance of the 
($\alpha$-dependent) ensemble Hx functional in this regime.
For $2 \leq \mathcal{N} \leq 3$,
the two-electron correlation energy
is not equal to zero such
that EEXX and GS-Hx differ slightly from
GS-c and GS-Hxc, respectively. Hence, GS-c is not
exact anymore, in contrast to EEXX.

As shown in the lower panels of Fig.~\ref{fig:asymmetric}, the curvature
of the approximate energies becomes more pronounced as $U/t$ increases. 
As expected, EEXX is not accurate anymore,
especially around $\mathcal{N}=2$, where the
correlation energy becomes important.
Unlike in the strongly-correlated symmetric case, GS-c
reproduces essentially the exact energy when $\mathcal{N} = 1$ (i.e.
$\alpha = 1$). In this case, the left $N$-centered
ensemble density is actually
equal to $n ^{\{N=2,\alpha=1\}}_- = 1 + \Delta v_{\rm ext}/\sqrt{4t^2 + \Delta v_{\rm ext}^2}$ [see Eq.~(\ref{eq:nleft})],
thus leading to
\begin{eqnarray}
n ^{\{N=2,\alpha=1\}}_- \xrightarrow{|\Delta v_{\rm ext}| /t \rightarrow +\infty}
2.
\end{eqnarray}
Similarly, it can be shown that, for $\alpha=1$, the right $N$-centered ensemble
density reads $n ^{\{N=2,\alpha=1\}}_+ = 1 + \Delta v_{\rm ext}/\left(3\sqrt{4t^2 + \Delta v_{\rm ext}^2}\right)$
such that
\begin{eqnarray}
n ^{\{N=2,\alpha=1\}}_+ \xrightarrow{|\Delta v_{\rm ext}| /t \rightarrow +\infty}
2 - \dfrac{2}{3}.
\end{eqnarray}
Thus we conclude that, for large $\Delta v_{\rm ext} /t$ values, the two ensemble densities reach the
border of their non-interacting $v$-representability domains
[see Eqs.~(\ref{eq:vKS-}) and (\ref{eq:vKS+})], such that
the exact 
($\alpha$-dependent) ensemble density-functional correlation energy
becomes zero. For the left ensemble, it is clear that
the
$\alpha$-independent correlation energy used in the GS-c approximation
is also equal to zero, as the border of the $v$-representability
domain does not depend on $\alpha$. Therefore, both GS-c and EEXX become
exact when $\mathcal{N}=1$. However, when $\mathcal{N}=3$, GS-c
introduces a spurious correlation energy contribution obtained by inserting the
right ensemble density value $n = 2-2/3$ into the two-electron
($\alpha=0$)
ground-state correlation functional. 

\section{Conclusions and perspectives}\label{sec:conclusions}

So-called left and right variants of $N$-centered ensemble DFT have been
explored. Unlike in the original formulation of the 
theory~\cite{senjean2018unified}, both open
and closed electronic systems can be described. In left/right
$N$-centered ensemble DFT, the key variables are 
the left/right ensemble density, which integrates to the central
integral number $N$ of electrons, {\it and} the (absolute) deviation $\alpha$ of
the true electron number from $N$. Within such a formalism, the
infamous derivative discontinuity is taken care of by the $\alpha$-dependent
$N$-centered ensemble Hxc functional. Its
$\alpha$-dependence plays also a key role in reproducing 
the correct piecewise linearity of the
energy, as illustrated in this work with the Hubbard dimer model.
What we learn from this model is that conventional
($\alpha$-independent) xc functionals are not sufficient for describing
open systems in the context of $N$-centered ensemble DFT. Developing
{\it ab initio} $\alpha$-dependent density-functional approximations is
a challenging but necessary task. As a starting point, a (semi-)local approximation could be obtained, for
example, by applying the theory to finite uniform electron gases~\cite{loos2017exchange}. Work is
currently in progress in this direction.\\

Various applications of left/right $N$-centered ensemble DFT can already
be foreseen. Regarding density-functional embedding techniques, it could
be used in place of conventional
DFT for grand canonical ensembles when describing open
fragments. The practical advantage would come from the explicit
description of derivative discontinuity contributions to the energy
through the dependence in $\alpha$ of the Hxc functional. Another even more appealing feature
of the $N$-centered formalism is the possibility to use a (pure-state)
$N$-electron
many-body wavefunction for describing an open fragment. Even though
such an idea seems counterintuitive and difficult to implement, it can
be made formally exact if appropriate density-functional corrections are
introduced. The basic idea is to consider the following decomposition: 

\be\label{eq:decomp_embedding}
F^{\{N,\alpha\}}_\pm\left[n_{\hat{\Gamma}_\pm^{\{N,\alpha\}}}\right]=F\left[n_{\hat{\Gamma}_\pm^{\{N,\alpha\}}}\right]+\Delta
F^{\{N,\alpha\}}_\pm\left[n_{\hat{\Gamma}_\pm^{\{N,\alpha\}}}\right],
\nonumber\\
\ee
where $n_{\hat{\Gamma}_\pm^{\{N,\alpha\}}}$ is the left/right $N$-centered
ensemble density [from which the density of the open system can be
extracted, according to Eqs.~(\ref{eq:true_dens_from_left}) and
(\ref{eq:true_dens_from_right})], and 
\be
\Delta
F^{\{N,\alpha\}}_\pm\left[n\right]=F^{\{N,\alpha\}}_\pm\left[n\right]-F^{\{N,\alpha=0\}}_\pm\left[n\right].
\ee
Since $F[n]=F^{\{N,\alpha=0\}}_\pm\left[n\right]$
is the standard Levy's
constrained-search functional, the first term on the right-hand
side of Eq.~(\ref{eq:decomp_embedding}) can be rewritten as 
\be\label{eq:pure-state_N-centered_dens}
F\left[n_{\hat{\Gamma}_\pm^{\{N,   \alpha\}}}\right]
&=&\min_{\Psi\rightarrow n_{\hat{\Gamma}_\pm^{\{N,\alpha\}}}}
\bra{\Psi}\hat{T}+\hat{W}_{\rm ee}\ket{\Psi}
\nonumber\\
&=&
\bra{\Psi_\pm^{\{N,\alpha\}}}\hat{T}+\hat{W}_{\rm ee}\ket{\Psi_\pm^{\{N,\alpha\}}}
,
\ee 
where $\Psi_\pm^{\{N,\alpha\}}$ is an auxiliary many-body wavefunction with density
$n_{\hat{\Gamma}_\pm^{\{N,\alpha\}}}$. Combining
Eqs.~(\ref{eq:var_princ_L-R_dens}), (\ref{eq:decomp_embedding}), and
(\ref{eq:pure-state_N-centered_dens}) leads to the following hybrid
wavefunction/DFT $N$-centered ensemble energy expression,    
\be
\mathscr{E}_\pm^{\{N,\alpha\}}&=&
\bra{\Psi_\pm^{\{N,\alpha\}}}\hat{T}+\hat{W}_{\rm ee}+\hat{V}_{\rm ext}\ket{\Psi_\pm^{\{N,\alpha\}}}
\nonumber\\
&&+\Delta
F^{\{N,\alpha\}}_\pm\left[n_{\Psi_\pm^{\{N,\alpha\}}}\right],
\ee
from which the energy of an open fragment can be extracted, in principle
exactly [see
Eqs.~(\ref{eq:true_ener_from_left})
and (\ref{eq:true_ener_from_right})]. Further exploration of this
formalism is left for future work. 

\section{Acknowledgments}
E.F. would like to thank Pierre-Fran{\c c}ois Loos for stimulating discussions about
ensembles and DFT.


\appendix

\section{Exact analytical expressions for the Hubbard dimer}\label{app:derivations}

In the following, the central number of
electrons will be set to $N = 2$.
The ensemble energies read
\be\label{eq:Eens_dimer}
\mathcal{E}_\pm^{\{N=2,\alpha\}}\left(\Delta v_{\rm
ext}\right)=\min_{n}\Bigg\{
&&T^{\{N=2,\alpha\}}_{\rm s\pm}(n)+E^{\{N=2,\alpha\}}_{\rm Hxc\pm}(n)
\nonumber\\
&&+\Delta v_{\rm
ext}\times\left(1-n\right)
\Bigg\},
\ee
where the minimizing densities are $n^{\{N=2,\alpha\}}_\pm\left(\Delta v_{\rm
ext}\right)$. Note that
\be\label{eq:Eens_derive_dimer}
&&\dfrac{\partial}{\partial\alpha} \mathcal{E}_\pm^{\{N=2,\alpha\}}\left(\Delta v_{\rm
ext}\right)
\nonumber\\
&&=\left[\dfrac{\partial T^{\{N=2,\alpha\}}_{\rm s\pm}(n)}{\partial\alpha}
+\dfrac{\partial E^{\{N=2,\alpha\}}_{\rm Hxc\pm}(n)}{\partial\alpha}
\right]_{n=n^{N=2,\alpha}_\pm\left(\Delta v_{\rm
ext}\right)},
\ee
where the Hx and the non-interacting kinetic density functionals
are given in Eqs.~(\ref{eq:Ts-}), (\ref{eq:Hx-}), 
(\ref{eq:Ts+}) and (\ref{eq:Hx+}), such that
\be
\dfrac{\partial T^{\{N=2,\alpha\}}_{\rm s-}(n)}{\partial \alpha}
=0,
\ee
\be
\dfrac{\partial E^{\{N=2,\alpha\}}_{\rm Hx-}(n)}{\partial \alpha}
=- \dfrac{U}{2}\left(1 + (1-n)^2\right),
\ee
\be
\dfrac{\partial T^{\{N=2,\alpha\}}_{\rm s+}(n)}{\partial \alpha}
=-\dfrac{4t}{9}\dfrac{(2\alpha-3)}{\sqrt{\left(\dfrac{2\alpha}{3} - 1\right)^2 -  (n - 1)^2}},
\ee
and
\be
\dfrac{\partial E^{\{N=2,\alpha\}}_{\rm Hx+}(n)}{\partial \alpha}
=\dfrac{U}{2}\left[\dfrac{1}{3}+\dfrac{9(2\alpha-1)(n-1)^2}{(2\alpha-3)^3}\right].
\ee
Plugging
Eqs.~(\ref{eq:Eens_dimer}) and (\ref{eq:Eens_derive_dimer}) 
into
Eqs.~(\ref{eq:true_ener_from_left})
and (\ref{eq:true_ener_from_right}),
one recovers the exact analytical expression
for the true physical energies in
Eq.~(\ref{eq:opensys_energy_Hubbard}).
In this article
we choose to focus
on the functional-driven errors only,
such that the exact left and right $N$-centered
ensemble densities
are used.
The latter are
obtained by differentiating the
left and right $N$-centered ensemble energies,
\be\label{eq:ens_en_-}
\mathscr{E}_-^{\{N=2,\alpha\}}\left(\Delta v_{\rm
ext}\right)=(1-\alpha)E^{N=2}\left(\Delta v_{\rm
ext}\right)
+
2\alpha \varepsilon_{\rm H}\left(\Delta v_{\rm 
ext}\right)
\nonumber\\
\ee
and
\be\label{eq:ens_en_+}
\mathscr{E}_+^{\{N=2,\alpha\}}\left(\Delta v_{\rm
ext}\right)&=&(1-\alpha)E^{N=2}\left(\Delta v_{\rm
ext}\right)
\nonumber\\
&&+
\dfrac{2\alpha}{3}\Big(\varepsilon_{\rm H}\left(\Delta v_{\rm 
ext}\right)+U\Big),
\ee
with respect to $\Delta v_{\rm ext}$, where
$\varepsilon_{\rm H}(\Delta v_{\rm ext}) = -(1/2)\sqrt{4t^2 + \Delta v_{\rm ext}^2}$ is the
HOMO energy.
According to the Hellmann--Feynman theorem,
\be
\dfrac{\partial \mathscr{E}_\pm^{\{N=2,\alpha\}}\left(\Delta v_{\rm
ext}\right)}{\partial \Delta v_{\rm
ext}}&=&
\dfrac{1}{2}{\rm Tr}\left[\hat{\Gamma}_\pm^{\{N=2,\alpha\}}\left(\hat{n}_1-\hat{n}_0\right)\right]
\nonumber\\
&=&
\dfrac{1}{2}{\rm Tr}\left[\hat{\Gamma}_\pm^{\{N=2,\alpha\}}\left(\hat{N}-2\hat{n}_0\right)\right]
\nonumber\\
&=&
\dfrac{N}{2}-n^{\{N=2,\alpha\}}_\pm\left(\Delta v_{\rm
ext}\right),
\ee
thus leading to the following expressions for the
left and right $N$-centered densities:
\be
n^{\{N=2,\alpha\}}_\pm\left(\Delta v_{\rm
ext}\right)=1-\dfrac{\partial \mathscr{E}_\pm^{\{N=2,\alpha\}}\left(\Delta v_{\rm
ext}\right)}{\partial \Delta v_{\rm ext}}.
\ee
Inserting Eqs.~(\ref{eq:ens_en_-}) and (\ref{eq:ens_en_+}) into
the latter expression leads to
\begin{eqnarray}\label{eq:nleft}
&&n^{\{N=2,\alpha\}}_-\left(\Delta v_{\rm
ext}\right) =
1 - 2\alpha \dfrac{\partial \varepsilon_{\rm H}(\Delta v_{\rm ext})}{\partial \Delta v_{\rm ext}}\nonumber\\
 &&- (1 - \alpha)\dfrac{\partial E^{N=2}(\Delta v_{\rm ext})}{\partial \Delta v_{\rm ext}},
\end{eqnarray}
and
\begin{eqnarray}\label{eq:nright}
&&n^{\{N=2,\alpha\}}_+\left(\Delta v_{\rm
ext}\right) =
1 - \dfrac{2\alpha}{3} \dfrac{\partial \varepsilon_{\rm H}(\Delta v_{\rm ext})}{\partial \Delta v_{\rm ext}}\nonumber\\
 &&- (1 - \alpha)\dfrac{\partial E^{N=2}(\Delta v_{\rm ext})}{\partial \Delta v_{\rm ext}},
\end{eqnarray}
where
\begin{eqnarray}
\dfrac{\partial \varepsilon_{\rm H}(\Delta v)}{\partial \Delta v}= - \dfrac{\Delta v}{2\sqrt{4t^2+\Delta v^2}},
\end{eqnarray}
The expressions for the
ground-state energy of the two-electron Hubbard dimer
and its derivatives are
given in the Appendix of Ref.~\cite{deur2017exact}.

\section{Piecewise linearity in the symmetric case}\label{app:symmetric}

In this appendix we derive the analytical formulas
for the approximate grand canonical energies
in the symmetric case ($\Delta v = 0$).
Let us start from
the expressions of the left and right
ensemble energies
\begin{eqnarray}
\mathscr{E}^{\{N=2,\alpha\}}_- (\Delta v)&=&
(1-\alpha) E^{N=2}(\Delta v) + 2\alpha E^{N=1} (\Delta v)\nonumber \\
\\
\mathscr{E}^{\{N=2,\alpha\}}_+(\Delta v)&=&
(1-\alpha) E^{N=2}(\Delta v) + 
\dfrac{2\alpha}{3} E^{N=3}(\Delta v),\nonumber \\
\end{eqnarray}
where $E^{N=1}(\Delta v = 0) = -t$, $E^{N=3}(\Delta v = 0) = U - t$
and
\begin{eqnarray}\label{eq:GS_ener_n1}
E^{N=2}(\Delta v = 0) = \dfrac{1}{2}\left(U - \sqrt{U^2 + 16t^2}\right).
\end{eqnarray}
Note that $\mathscr{E}^{\{N,\alpha\}}_\pm (\Delta v = 0) = 
F_{\pm}^{\{N,\alpha\}}(n=1)$ thus leading to,
\begin{eqnarray}
F_{-}^{\{N=2,\alpha\}}(n=1) & =  & \dfrac{1-\alpha}{2} \left(
U - \sqrt{U^2 + 16t^2}\right)
-2\alpha t, \\
F_{+}^{\{N=2,\alpha\}}(n=1) & = &  \dfrac{1-\alpha}{2} \left(
U - \sqrt{U^2 + 16t^2}\right) + \dfrac{2\alpha}{3} (U-t).\nonumber\\
\end{eqnarray}
As the non-interacting kinetic energy functionals read
\begin{eqnarray}
T_{\rm s-}^{\{N=2,\alpha\}}(n=1) &=& -2t,\\
T_{\rm s+}^{\{N=2,\alpha\}}(n=1) & = & -2t\left( 
1 - \dfrac{2\alpha}{3}\right),
\end{eqnarray}
one obtains the following
expression for the left and right Hxc energy functionals:
\begin{eqnarray}
E_{\rm Hxc-}^{\{N=2, \alpha\}}(n=1)
&=&  \dfrac{1-\alpha}{2} \left(
U - \sqrt{U^2 + 16t^2} + 4t\right),\\
E_{\rm Hxc+}^{\{N=2, \alpha\}}(n=1)
&=& \dfrac{2}{3}\alpha U + E_{\rm Hxc-}^{\{N=2, \alpha\}}(n=1).
\end{eqnarray}
The grand canonical energies
are then given by
\begin{eqnarray}
E(\mathcal{N}=2-\alpha) &=& \dfrac{(2-\alpha)}{2}\left( E_{\rm Hxc-}^{\{N,\alpha\}}(n=1) - 2t\right) \nonumber \\
&&- \dfrac{\alpha(1-\alpha)}{2} \dfrac{\partial E_{\rm Hxc-}^{\{N,\alpha\}}(n=1)}{\partial \alpha} \label{eq:EN-alpha-sym}\\
E(\mathcal{N}=2+\alpha) &=& -2t\left(1-\dfrac{\alpha}{2}\right) 
+ \left(1+\dfrac{\alpha}{2}\right) E_{\rm Hxc+}^{\{N,\alpha\}}(n=1) \nonumber \\
&&+ \dfrac{\alpha(1-\alpha)}{2} \dfrac{\partial E_{\rm Hxc+}^{\{N,\alpha\}}(n=1)}{\partial \alpha} \label{eq:EN+alpha-sym}
\end{eqnarray}
By realizing that
\begin{eqnarray}
E_{\rm Hxc-}^{\{N=2, \alpha\}}(n=1)
&=& (\alpha - 1)\dfrac{\partial E_{\rm Hxc-}^{\{N=2, \alpha\}}(n=1)}{\partial \alpha}, \\
E_{\rm Hxc+}^{\{N=2, \alpha\}}(n=1)
&=& (\alpha - 1) \dfrac{\partial E_{\rm Hxc+}^{\{N=2, \alpha\}}(n=1)}{\partial \alpha} + \dfrac{2}{3}U, \nonumber \\
\end{eqnarray}
where $\partial E_{\rm Hxc\pm}^{\{N=2, \alpha\}}(n=1)/ \partial \alpha$
is constant ($\alpha$-independent), it is straightforward to
see that all quadratic terms (with respect to $\alpha$)
in Eqs.~(\ref{eq:EN-alpha-sym}) and (\ref{eq:EN+alpha-sym}) 
cancel out as expected.
The slopes of the true physical energies
are, according to Eqs.~(\ref{eq:gamma_matchalN}) and (\ref{eq:GS_ener_n1}),
\begin{eqnarray}\label{eq:slope_exact}
\dfrac{\partial E(\mathcal{N}=2\pm\alpha)}{\partial \alpha} &=& \dfrac{1}{2}\left(\pm U - 2t + \sqrt{U^2 + 16t^2}\right).
\end{eqnarray}
Turning to the different approximations considered in 
Sec.~\ref{sec:hubbard} of this paper,
we follow the same reasoning
and end up with the following constant slopes for $\Delta v = 0$
for GS-Hx,
\begin{eqnarray}
\dfrac{\partial E(\mathcal{N}=2\pm\alpha)}{\partial \alpha} = t \pm \dfrac{U}{4},
\end{eqnarray}
for EEXX,
\begin{eqnarray}
\dfrac{\partial E(\mathcal{N}=2\pm\alpha)}{\partial \alpha} = t \pm \dfrac{U}{2},
\end{eqnarray}
for GS-Hxc,
\begin{eqnarray}
\dfrac{\partial E(\mathcal{N}=2-\alpha)}{\partial \alpha} &=&
- \dfrac{1}{4}\left(
U - \sqrt{U^2 + 16t^2}\right),\label{eq:slope_GS-Hxc}
\\
 \dfrac{\partial E(\mathcal{N}=2+\alpha)}{\partial \alpha} &=&
2t + \dfrac{1}{4}\left(
U - \sqrt{U^2 + 16t^2}\right),
\end{eqnarray}
and finally for GS-c,
\begin{eqnarray}
\dfrac{\partial E(\mathcal{N}=2-\alpha)}{\partial \alpha} &=&
- \dfrac{U}{2} + \dfrac{1}{4}\sqrt{U^2 + 16t^2},
\\
 \dfrac{\partial E(\mathcal{N}=2+\alpha)}{\partial \alpha} &=&
2t + \dfrac{U}{2} - \dfrac{1}{4} \sqrt{U^2 + 16t^2}.
\end{eqnarray}
Note that in the case $\mathcal{N} = 2 - \alpha$, the slope
with respect to $\mathcal{N}$ in
Figs.~\ref{fig:symmetric} and \ref{fig:asymmetric}
are the same as the above but with opposite sign.
As readily seen from
Eqs.~(\ref{eq:slope_exact}) and (\ref{eq:slope_GS-Hxc})
in the atomic limit $t \rightarrow 0$,
the energy of the open system
within GS-Hxc has the same slope (equal to zero)
as
the exact one for $1 \leq \mathcal{N} \leq 2$. 
As GS-Hxc is
exact at $\mathcal{N}=2$, GS-Hxc is also exact in the range
$1 \leq \mathcal{N} \leq 2$ in the atomic limit.
Another interesting behaviour, in the symmetric case and
in the atomic limit,
is that none of the approximations
considered in this article can describe the derivative discontinuity
of the grand canonical energy when crossing $\mathcal{N} = N = 2$.
Indeed, all the approximations feature a constant slope
in between $1 \leq \mathcal{N} \leq 3$, making no difference between the ionization potential
and the electronic affinity, and thus no fundamental gap.
This is obviously not correct, as the exact fundamental gap is equal to $U$
in this limit. Therefore, in order
to reproduce the correct behaviour, weight-dependent
correlation functionals are required.
Such developments are
left for future work.

\newcommand{\Aa}[0]{Aa}



\begin{thebibliography}{55}%
\makeatletter
\providecommand \@ifxundefined [1]{%
 \@ifx{#1\undefined}
}%
\providecommand \@ifnum [1]{%
 \ifnum #1\expandafter \@firstoftwo
 \else \expandafter \@secondoftwo
 \fi
}%
\providecommand \@ifx [1]{%
 \ifx #1\expandafter \@firstoftwo
 \else \expandafter \@secondoftwo
 \fi
}%
\providecommand \natexlab [1]{#1}%
\providecommand \enquote  [1]{``#1''}%
\providecommand \bibnamefont  [1]{#1}%
\providecommand \bibfnamefont [1]{#1}%
\providecommand \citenamefont [1]{#1}%
\providecommand \href@noop [0]{\@secondoftwo}%
\providecommand \href [0]{\begingroup \@sanitize@url \@href}%
\providecommand \@href[1]{\@@startlink{#1}\@@href}%
\providecommand \@@href[1]{\endgroup#1\@@endlink}%
\providecommand \@sanitize@url [0]{\catcode `\\12\catcode `\$12\catcode
  `\&12\catcode `\#12\catcode `\^12\catcode `\_12\catcode `\%12\relax}%
\providecommand \@@startlink[1]{}%
\providecommand \@@endlink[0]{}%
\providecommand \url  [0]{\begingroup\@sanitize@url \@url }%
\providecommand \@url [1]{\endgroup\@href {#1}{\urlprefix }}%
\providecommand \urlprefix  [0]{URL }%
\providecommand \Eprint [0]{\href }%
\providecommand \doibase [0]{http://dx.doi.org/}%
\providecommand \selectlanguage [0]{\@gobble}%
\providecommand \bibinfo  [0]{\@secondoftwo}%
\providecommand \bibfield  [0]{\@secondoftwo}%
\providecommand \translation [1]{[#1]}%
\providecommand \BibitemOpen [0]{}%
\providecommand \bibitemStop [0]{}%
\providecommand \bibitemNoStop [0]{.\EOS\space}%
\providecommand \EOS [0]{\spacefactor3000\relax}%
\providecommand \BibitemShut  [1]{\csname bibitem#1\endcsname}%
\let\auto@bib@innerbib\@empty
\bibitem [{\citenamefont {Perdew}\ \emph {et~al.}(1982)\citenamefont {Perdew},
  \citenamefont {Parr}, \citenamefont {Levy},\ and\ \citenamefont
  {Balduz~Jr}}]{perdew1982density}%
  \BibitemOpen
  \bibfield  {author} {\bibinfo {author} {\bibfnamefont {J.~P.}\ \bibnamefont
  {Perdew}}, \bibinfo {author} {\bibfnamefont {R.~G.}\ \bibnamefont {Parr}},
  \bibinfo {author} {\bibfnamefont {M.}~\bibnamefont {Levy}}, \ and\ \bibinfo
  {author} {\bibfnamefont {J.~L.}\ \bibnamefont {Balduz~Jr}},\ }\href
  {https://doi.org/10.1103/PhysRevLett.49.1691} {\bibfield  {journal} {\bibinfo
   {journal} {Phys. Rev. Lett.}\ }\textbf {\bibinfo {volume} {49}},\ \bibinfo
  {pages} {1691} (\bibinfo {year} {1982})}\BibitemShut {NoStop}%
\bibitem [{\citenamefont {Perdew}\ and\ \citenamefont
  {Levy}(1983)}]{perdew1983physical}%
  \BibitemOpen
  \bibfield  {author} {\bibinfo {author} {\bibfnamefont {J.~P.}\ \bibnamefont
  {Perdew}}\ and\ \bibinfo {author} {\bibfnamefont {M.}~\bibnamefont {Levy}},\
  }\href {https://doi.org/10.1103/PhysRevLett.51.1884} {\bibfield  {journal}
  {\bibinfo  {journal} {Phys. Rev. Lett.}\ }\textbf {\bibinfo {volume} {51}},\
  \bibinfo {pages} {1884} (\bibinfo {year} {1983})}\BibitemShut {NoStop}%
\bibitem [{\citenamefont {Cohen}\ and\ \citenamefont
  {Wasserman}(2006)}]{cohen2006hardness}%
  \BibitemOpen
  \bibfield  {author} {\bibinfo {author} {\bibfnamefont {M.~H.}\ \bibnamefont
  {Cohen}}\ and\ \bibinfo {author} {\bibfnamefont {A.}~\bibnamefont
  {Wasserman}},\ }\href {https://doi.org/10.1007/s10955-006-9031-0} {\bibfield
  {journal} {\bibinfo  {journal} {J. Stat. Phys.}\ }\textbf {\bibinfo {volume}
  {125}},\ \bibinfo {pages} {1121} (\bibinfo {year} {2006})}\BibitemShut
  {NoStop}%
\bibitem [{\citenamefont {Cohen}\ and\ \citenamefont
  {Wasserman}(2007)}]{cohen2007foundations}%
  \BibitemOpen
  \bibfield  {author} {\bibinfo {author} {\bibfnamefont {M.~H.}\ \bibnamefont
  {Cohen}}\ and\ \bibinfo {author} {\bibfnamefont {A.}~\bibnamefont
  {Wasserman}},\ }\href {https://doi.org/10.1021/jp066449h} {\bibfield
  {journal} {\bibinfo  {journal} {J. Phys. Chem. A}\ }\textbf {\bibinfo
  {volume} {111}},\ \bibinfo {pages} {2229} (\bibinfo {year}
  {2007})}\BibitemShut {NoStop}%
\bibitem [{\citenamefont {Elliott}\ \emph {et~al.}(2009)\citenamefont
  {Elliott}, \citenamefont {Cohen}, \citenamefont {Wasserman},\ and\
  \citenamefont {Burke}}]{elliott2009density}%
  \BibitemOpen
  \bibfield  {author} {\bibinfo {author} {\bibfnamefont {P.}~\bibnamefont
  {Elliott}}, \bibinfo {author} {\bibfnamefont {M.~H.}\ \bibnamefont {Cohen}},
  \bibinfo {author} {\bibfnamefont {A.}~\bibnamefont {Wasserman}}, \ and\
  \bibinfo {author} {\bibfnamefont {K.}~\bibnamefont {Burke}},\ }\href
  {https://doi.org/10.1021/ct9000119} {\bibfield  {journal} {\bibinfo
  {journal} {J. Chem. Theory Comput.}\ }\textbf {\bibinfo {volume} {5}},\
  \bibinfo {pages} {827} (\bibinfo {year} {2009})}\BibitemShut {NoStop}%
\bibitem [{\citenamefont {Elliott}\ \emph {et~al.}(2010)\citenamefont
  {Elliott}, \citenamefont {Burke}, \citenamefont {Cohen},\ and\ \citenamefont
  {Wasserman}}]{elliott2010partition}%
  \BibitemOpen
  \bibfield  {author} {\bibinfo {author} {\bibfnamefont {P.}~\bibnamefont
  {Elliott}}, \bibinfo {author} {\bibfnamefont {K.}~\bibnamefont {Burke}},
  \bibinfo {author} {\bibfnamefont {M.~H.}\ \bibnamefont {Cohen}}, \ and\
  \bibinfo {author} {\bibfnamefont {A.}~\bibnamefont {Wasserman}},\ }\href
  {https://doi.org/10.1103/PhysRevA.82.024501} {\bibfield  {journal} {\bibinfo
  {journal} {Phys. Rev. A}\ }\textbf {\bibinfo {volume} {82}},\ \bibinfo
  {pages} {024501} (\bibinfo {year} {2010})}\BibitemShut {NoStop}%
\bibitem [{\citenamefont {Nafziger}\ \emph {et~al.}(2011)\citenamefont
  {Nafziger}, \citenamefont {Wu},\ and\ \citenamefont
  {Wasserman}}]{nafziger2011molecular}%
  \BibitemOpen
  \bibfield  {author} {\bibinfo {author} {\bibfnamefont {J.}~\bibnamefont
  {Nafziger}}, \bibinfo {author} {\bibfnamefont {Q.}~\bibnamefont {Wu}}, \ and\
  \bibinfo {author} {\bibfnamefont {A.}~\bibnamefont {Wasserman}},\ }\href
  {https://doi.org/10.1063/1.3667198} {\bibfield  {journal} {\bibinfo
  {journal} {J. Chem. Phys.}\ }\textbf {\bibinfo {volume} {135}},\ \bibinfo
  {pages} {234101} (\bibinfo {year} {2011})}\BibitemShut {NoStop}%
\bibitem [{\citenamefont {Tang}\ \emph {et~al.}(2012)\citenamefont {Tang},
  \citenamefont {Nafziger},\ and\ \citenamefont
  {Wasserman}}]{tang2012fragment}%
  \BibitemOpen
  \bibfield  {author} {\bibinfo {author} {\bibfnamefont {R.}~\bibnamefont
  {Tang}}, \bibinfo {author} {\bibfnamefont {J.}~\bibnamefont {Nafziger}}, \
  and\ \bibinfo {author} {\bibfnamefont {A.}~\bibnamefont {Wasserman}},\ }\href
  {http://dx.doi.org/10.1039/C2CP23994A} {\bibfield  {journal} {\bibinfo
  {journal} {Phys. Chem. Chem. Phys.}\ }\textbf {\bibinfo {volume} {14}},\
  \bibinfo {pages} {7780} (\bibinfo {year} {2012})}\BibitemShut {NoStop}%
\bibitem [{\citenamefont {Mosquera}\ and\ \citenamefont
  {Wasserman}(2013)}]{mosquera2013partition}%
  \BibitemOpen
  \bibfield  {author} {\bibinfo {author} {\bibfnamefont {M.~A.}\ \bibnamefont
  {Mosquera}}\ and\ \bibinfo {author} {\bibfnamefont {A.}~\bibnamefont
  {Wasserman}},\ }\href {https://doi.org/10.1080/00268976.2012.729096}
  {\bibfield  {journal} {\bibinfo  {journal} {Mol. Phys.}\ }\textbf {\bibinfo
  {volume} {111}},\ \bibinfo {pages} {505} (\bibinfo {year}
  {2013})}\BibitemShut {NoStop}%
\bibitem [{\citenamefont {Niffenegger}\ \emph {et~al.}(2019)\citenamefont
  {Niffenegger}, \citenamefont {Oueis}, \citenamefont {Nafziger},\ and\
  \citenamefont {Wasserman}}]{niffenegger2019density}%
  \BibitemOpen
  \bibfield  {author} {\bibinfo {author} {\bibfnamefont {K.}~\bibnamefont
  {Niffenegger}}, \bibinfo {author} {\bibfnamefont {Y.}~\bibnamefont {Oueis}},
  \bibinfo {author} {\bibfnamefont {J.}~\bibnamefont {Nafziger}}, \ and\
  \bibinfo {author} {\bibfnamefont {A.}~\bibnamefont {Wasserman}},\ }\href
  {https://doi.org/10.1080/00268976.2019.1618939} {\bibfield  {journal}
  {\bibinfo  {journal} {Mol. Phys.}\ ,\ \bibinfo {pages} {1}} (\bibinfo {year}
  {2019})}\BibitemShut {NoStop}%
\bibitem [{\citenamefont {Huang}\ and\ \citenamefont
  {Carter}(2011)}]{huang2011potential}%
  \BibitemOpen
  \bibfield  {author} {\bibinfo {author} {\bibfnamefont {C.}~\bibnamefont
  {Huang}}\ and\ \bibinfo {author} {\bibfnamefont {E.~A.}\ \bibnamefont
  {Carter}},\ }\href {\doibase 10.1063/1.3659293} {\bibfield  {journal}
  {\bibinfo  {journal} {J. Chem. Phys.}\ }\textbf {\bibinfo {volume} {135}},\
  \bibinfo {pages} {194104} (\bibinfo {year} {2011})}\BibitemShut {NoStop}%
\bibitem [{\citenamefont {Fabiano}\ \emph {et~al.}(2014)\citenamefont
  {Fabiano}, \citenamefont {Laricchia},\ and\ \citenamefont
  {Sala}}]{fabiano2014frozen}%
  \BibitemOpen
  \bibfield  {author} {\bibinfo {author} {\bibfnamefont {E.}~\bibnamefont
  {Fabiano}}, \bibinfo {author} {\bibfnamefont {S.}~\bibnamefont {Laricchia}},
  \ and\ \bibinfo {author} {\bibfnamefont {F.~D.}\ \bibnamefont {Sala}},\
  }\href {\doibase 10.1063/1.4868033} {\bibfield  {journal} {\bibinfo
  {journal} {J. Chem. Phys.}\ }\textbf {\bibinfo {volume} {140}},\ \bibinfo
  {pages} {114101} (\bibinfo {year} {2014})}\BibitemShut {NoStop}%
\bibitem [{\citenamefont {Kraisler}\ and\ \citenamefont
  {Kronik}(2013)}]{kraisler2013piecewise}%
  \BibitemOpen
  \bibfield  {author} {\bibinfo {author} {\bibfnamefont {E.}~\bibnamefont
  {Kraisler}}\ and\ \bibinfo {author} {\bibfnamefont {L.}~\bibnamefont
  {Kronik}},\ }\href {https://doi.org/10.1103/PhysRevLett.110.126403}
  {\bibfield  {journal} {\bibinfo  {journal} {Phys. Rev. Lett.}\ }\textbf
  {\bibinfo {volume} {110}},\ \bibinfo {pages} {126403} (\bibinfo {year}
  {2013})}\BibitemShut {NoStop}%
\bibitem [{\citenamefont {Senjean}\ and\ \citenamefont
  {Fromager}(2018)}]{senjean2018unified}%
  \BibitemOpen
  \bibfield  {author} {\bibinfo {author} {\bibfnamefont {B.}~\bibnamefont
  {Senjean}}\ and\ \bibinfo {author} {\bibfnamefont {E.}~\bibnamefont
  {Fromager}},\ }\href {https://doi.org/10.1103/PhysRevA.98.022513} {\bibfield
  {journal} {\bibinfo  {journal} {Phys. Rev. A}\ }\textbf {\bibinfo {volume}
  {98}},\ \bibinfo {pages} {022513} (\bibinfo {year} {2018})}\BibitemShut
  {NoStop}%
\bibitem [{\citenamefont {Gross}\ \emph {et~al.}(1988)\citenamefont {Gross},
  \citenamefont {Oliveira},\ and\ \citenamefont {Kohn}}]{gross1988density}%
  \BibitemOpen
  \bibfield  {author} {\bibinfo {author} {\bibfnamefont {E.~K.~U.}\
  \bibnamefont {Gross}}, \bibinfo {author} {\bibfnamefont {L.~N.}\ \bibnamefont
  {Oliveira}}, \ and\ \bibinfo {author} {\bibfnamefont {W.}~\bibnamefont
  {Kohn}},\ }\href {https://doi.org/10.1103/PhysRevA.37.2809} {\bibfield
  {journal} {\bibinfo  {journal} {Phys. Rev. A}\ }\textbf {\bibinfo {volume}
  {37}},\ \bibinfo {pages} {2809} (\bibinfo {year} {1988})}\BibitemShut
  {NoStop}%
\bibitem [{\citenamefont {Gould}\ and\ \citenamefont
  {Pittalis}(2019)}]{gould2019density}%
  \BibitemOpen
  \bibfield  {author} {\bibinfo {author} {\bibfnamefont {T.}~\bibnamefont
  {Gould}}\ and\ \bibinfo {author} {\bibfnamefont {S.}~\bibnamefont
  {Pittalis}},\ }\href
  {https://link.aps.org/doi/10.1103/PhysRevLett.123.016401} {\bibfield
  {journal} {\bibinfo  {journal} {Phys. Rev. Lett.}\ }\textbf {\bibinfo
  {volume} {123}},\ \bibinfo {pages} {016401} (\bibinfo {year}
  {2019})}\BibitemShut {NoStop}%
\bibitem [{\citenamefont {Gould}\ and\ \citenamefont
  {Pittalis}(2020)}]{gould2020density}%
  \BibitemOpen
  \bibfield  {author} {\bibinfo {author} {\bibfnamefont {T.}~\bibnamefont
  {Gould}}\ and\ \bibinfo {author} {\bibfnamefont {S.}~\bibnamefont
  {Pittalis}},\ }\href {https://arxiv.org/abs/2001.09429} {\bibfield  {journal}
  {\bibinfo  {journal} {arXiv:2001.09429}\ } (\bibinfo {year}
  {2020})}\BibitemShut {NoStop}%
\bibitem [{\citenamefont {Fromager}(2020)}]{fromager2020individual}%
  \BibitemOpen
  \bibfield  {author} {\bibinfo {author} {\bibfnamefont {E.}~\bibnamefont
  {Fromager}},\ }\href {https://arxiv.org/abs/2001.08605} {\bibfield  {journal}
  {\bibinfo  {journal} {arXiv:2001.08605}\ } (\bibinfo {year}
  {2020})}\BibitemShut {NoStop}%
\bibitem [{\citenamefont {Baerends}(2017)}]{baerends2017kohn}%
  \BibitemOpen
  \bibfield  {author} {\bibinfo {author} {\bibfnamefont {E.~J.}\ \bibnamefont
  {Baerends}},\ }\href {http://dx.doi.org/10.1039/C7CP02123B} {\bibfield
  {journal} {\bibinfo  {journal} {Phys. Chem. Chem. Phys.}\ }\textbf {\bibinfo
  {volume} {19}},\ \bibinfo {pages} {15639} (\bibinfo {year}
  {2017})}\BibitemShut {NoStop}%
\bibitem [{\citenamefont {Andrade}\ and\ \citenamefont
  {Aspuru-Guzik}(2011)}]{andrade2011prediction}%
  \BibitemOpen
  \bibfield  {author} {\bibinfo {author} {\bibfnamefont {X.}~\bibnamefont
  {Andrade}}\ and\ \bibinfo {author} {\bibfnamefont {A.}~\bibnamefont
  {Aspuru-Guzik}},\ }\href {https://doi.org/10.1103/PhysRevLett.107.183002}
  {\bibfield  {journal} {\bibinfo  {journal} {Phys. Rev. Lett.}\ }\textbf
  {\bibinfo {volume} {107}},\ \bibinfo {pages} {183002} (\bibinfo {year}
  {2011})}\BibitemShut {NoStop}%
\bibitem [{\citenamefont {Atalla}\ \emph {et~al.}(2016)\citenamefont {Atalla},
  \citenamefont {Zhang}, \citenamefont {Hofmann}, \citenamefont {Ren},
  \citenamefont {Rinke},\ and\ \citenamefont
  {Scheffler}}]{atalla2016enforcing}%
  \BibitemOpen
  \bibfield  {author} {\bibinfo {author} {\bibfnamefont {V.}~\bibnamefont
  {Atalla}}, \bibinfo {author} {\bibfnamefont {I.~Y.}\ \bibnamefont {Zhang}},
  \bibinfo {author} {\bibfnamefont {O.~T.}\ \bibnamefont {Hofmann}}, \bibinfo
  {author} {\bibfnamefont {X.}~\bibnamefont {Ren}}, \bibinfo {author}
  {\bibfnamefont {P.}~\bibnamefont {Rinke}}, \ and\ \bibinfo {author}
  {\bibfnamefont {M.}~\bibnamefont {Scheffler}},\ }\href
  {https://doi.org/10.1103/PhysRevB.94.035140} {\bibfield  {journal} {\bibinfo
  {journal} {Phys. Rev. B}\ }\textbf {\bibinfo {volume} {94}},\ \bibinfo
  {pages} {035140} (\bibinfo {year} {2016})}\BibitemShut {NoStop}%
\bibitem [{\citenamefont {Bhandari}\ \emph {et~al.}(2018)\citenamefont
  {Bhandari}, \citenamefont {Cheung}, \citenamefont {Geva}, \citenamefont
  {Kronik},\ and\ \citenamefont {Dunietz}}]{bhandari2018fundamental}%
  \BibitemOpen
  \bibfield  {author} {\bibinfo {author} {\bibfnamefont {S.}~\bibnamefont
  {Bhandari}}, \bibinfo {author} {\bibfnamefont {M.~S.}\ \bibnamefont
  {Cheung}}, \bibinfo {author} {\bibfnamefont {E.}~\bibnamefont {Geva}},
  \bibinfo {author} {\bibfnamefont {L.}~\bibnamefont {Kronik}}, \ and\ \bibinfo
  {author} {\bibfnamefont {B.~D.}\ \bibnamefont {Dunietz}},\ }\href
  {https://doi.org/10.1021/acs.jctc.8b00876} {\bibfield  {journal} {\bibinfo
  {journal} {J. Chem. Theory Comput.}\ }\textbf {\bibinfo {volume} {14}},\
  \bibinfo {pages} {6287} (\bibinfo {year} {2018})}\BibitemShut {NoStop}%
\bibitem [{\citenamefont {Elmaslmane}\ \emph {et~al.}(2018)\citenamefont
  {Elmaslmane}, \citenamefont {Wetherell}, \citenamefont {Hodgson},
  \citenamefont {McKenna},\ and\ \citenamefont
  {Godby}}]{elmaslmane2018accuracy}%
  \BibitemOpen
  \bibfield  {author} {\bibinfo {author} {\bibfnamefont {A.~R.}\ \bibnamefont
  {Elmaslmane}}, \bibinfo {author} {\bibfnamefont {J.}~\bibnamefont
  {Wetherell}}, \bibinfo {author} {\bibfnamefont {M.~J.~P.}\ \bibnamefont
  {Hodgson}}, \bibinfo {author} {\bibfnamefont {K.~P.}\ \bibnamefont
  {McKenna}}, \ and\ \bibinfo {author} {\bibfnamefont {R.~W.}\ \bibnamefont
  {Godby}},\ }\href {https://doi.org/10.1103/PhysRevMaterials.2.040801}
  {\bibfield  {journal} {\bibinfo  {journal} {Phys. Rev. Materials}\ }\textbf
  {\bibinfo {volume} {2}},\ \bibinfo {pages} {040801} (\bibinfo {year}
  {2018})}\BibitemShut {NoStop}%
\bibitem [{\citenamefont {Imamura}\ \emph {et~al.}(2011)\citenamefont
  {Imamura}, \citenamefont {Kobayashi},\ and\ \citenamefont
  {Nakai}}]{imamura2011linearity}%
  \BibitemOpen
  \bibfield  {author} {\bibinfo {author} {\bibfnamefont {Y.}~\bibnamefont
  {Imamura}}, \bibinfo {author} {\bibfnamefont {R.}~\bibnamefont {Kobayashi}},
  \ and\ \bibinfo {author} {\bibfnamefont {H.}~\bibnamefont {Nakai}},\ }\href
  {https://doi.org/10.1063/1.3569030} {\bibfield  {journal} {\bibinfo
  {journal} {J. Chem. Phys.}\ }\textbf {\bibinfo {volume} {134}},\ \bibinfo
  {pages} {124113} (\bibinfo {year} {2011})}\BibitemShut {NoStop}%
\bibitem [{\citenamefont {Kronik}\ \emph {et~al.}(2012)\citenamefont {Kronik},
  \citenamefont {Stein}, \citenamefont {Refaely-Abramson},\ and\ \citenamefont
  {Baer}}]{kronik2012excitation}%
  \BibitemOpen
  \bibfield  {author} {\bibinfo {author} {\bibfnamefont {L.}~\bibnamefont
  {Kronik}}, \bibinfo {author} {\bibfnamefont {T.}~\bibnamefont {Stein}},
  \bibinfo {author} {\bibfnamefont {S.}~\bibnamefont {Refaely-Abramson}}, \
  and\ \bibinfo {author} {\bibfnamefont {R.}~\bibnamefont {Baer}},\ }\href
  {https://doi.org/10.1021/ct2009363} {\bibfield  {journal} {\bibinfo
  {journal} {J. Chem. Theory Comput.}\ }\textbf {\bibinfo {volume} {8}},\
  \bibinfo {pages} {1515} (\bibinfo {year} {2012})}\BibitemShut {NoStop}%
\bibitem [{\citenamefont {Zheng}\ \emph {et~al.}(2011)\citenamefont {Zheng},
  \citenamefont {Cohen}, \citenamefont {Mori-S{\'a}nchez}, \citenamefont {Hu},\
  and\ \citenamefont {Yang}}]{zheng2011improving}%
  \BibitemOpen
  \bibfield  {author} {\bibinfo {author} {\bibfnamefont {X.}~\bibnamefont
  {Zheng}}, \bibinfo {author} {\bibfnamefont {A.~J.}\ \bibnamefont {Cohen}},
  \bibinfo {author} {\bibfnamefont {P.}~\bibnamefont {Mori-S{\'a}nchez}},
  \bibinfo {author} {\bibfnamefont {X.}~\bibnamefont {Hu}}, \ and\ \bibinfo
  {author} {\bibfnamefont {W.}~\bibnamefont {Yang}},\ }\href
  {https://doi.org/10.1103/PhysRevLett.107.026403} {\bibfield  {journal}
  {\bibinfo  {journal} {Phys. Rev. Lett.}\ }\textbf {\bibinfo {volume} {107}},\
  \bibinfo {pages} {026403} (\bibinfo {year} {2011})}\BibitemShut {NoStop}%
\bibitem [{\citenamefont {Zheng}\ \emph {et~al.}(2013)\citenamefont {Zheng},
  \citenamefont {Zhou},\ and\ \citenamefont {Yang}}]{zheng2013nonempirical}%
  \BibitemOpen
  \bibfield  {author} {\bibinfo {author} {\bibfnamefont {X.}~\bibnamefont
  {Zheng}}, \bibinfo {author} {\bibfnamefont {T.}~\bibnamefont {Zhou}}, \ and\
  \bibinfo {author} {\bibfnamefont {W.}~\bibnamefont {Yang}},\ }\href
  {https://doi.org/10.1063/1.4801922} {\bibfield  {journal} {\bibinfo
  {journal} {J. Chem. Phys.}\ }\textbf {\bibinfo {volume} {138}},\ \bibinfo
  {pages} {174105} (\bibinfo {year} {2013})}\BibitemShut {NoStop}%
\bibitem [{\citenamefont {Li}\ \emph {et~al.}(2015)\citenamefont {Li},
  \citenamefont {Zheng}, \citenamefont {Cohen}, \citenamefont
  {Mori-S{\'a}nchez},\ and\ \citenamefont {Yang}}]{li2015local}%
  \BibitemOpen
  \bibfield  {author} {\bibinfo {author} {\bibfnamefont {C.}~\bibnamefont
  {Li}}, \bibinfo {author} {\bibfnamefont {X.}~\bibnamefont {Zheng}}, \bibinfo
  {author} {\bibfnamefont {A.~J.}\ \bibnamefont {Cohen}}, \bibinfo {author}
  {\bibfnamefont {P.}~\bibnamefont {Mori-S{\'a}nchez}}, \ and\ \bibinfo
  {author} {\bibfnamefont {W.}~\bibnamefont {Yang}},\ }\href
  {https://doi.org/10.1103/PhysRevLett.114.053001} {\bibfield  {journal}
  {\bibinfo  {journal} {Phys. Rev. Lett.}\ }\textbf {\bibinfo {volume} {114}},\
  \bibinfo {pages} {053001} (\bibinfo {year} {2015})}\BibitemShut {NoStop}%
\bibitem [{\citenamefont {Stein}\ \emph {et~al.}(2012)\citenamefont {Stein},
  \citenamefont {Autschbach}, \citenamefont {Govind}, \citenamefont {Kronik},\
  and\ \citenamefont {Baer}}]{stein2012curvature}%
  \BibitemOpen
  \bibfield  {author} {\bibinfo {author} {\bibfnamefont {T.}~\bibnamefont
  {Stein}}, \bibinfo {author} {\bibfnamefont {J.}~\bibnamefont {Autschbach}},
  \bibinfo {author} {\bibfnamefont {N.}~\bibnamefont {Govind}}, \bibinfo
  {author} {\bibfnamefont {L.}~\bibnamefont {Kronik}}, \ and\ \bibinfo {author}
  {\bibfnamefont {R.}~\bibnamefont {Baer}},\ }\href
  {https://doi.org/10.1021/jz3015937} {\bibfield  {journal} {\bibinfo
  {journal} {J. Chem. Phys. Lett.}\ }\textbf {\bibinfo {volume} {3}},\ \bibinfo
  {pages} {3740} (\bibinfo {year} {2012})}\BibitemShut {NoStop}%
\bibitem [{\citenamefont {Hait}\ and\ \citenamefont
  {Head-Gordon}(2018)}]{hait2018delocalization}%
  \BibitemOpen
  \bibfield  {author} {\bibinfo {author} {\bibfnamefont {D.}~\bibnamefont
  {Hait}}\ and\ \bibinfo {author} {\bibfnamefont {M.}~\bibnamefont
  {Head-Gordon}},\ }\href {https://doi.org/10.1021/acs.jpclett.8b02417}
  {\bibfield  {journal} {\bibinfo  {journal} {J. Phys. Chem. Lett.}\ }\textbf
  {\bibinfo {volume} {9}},\ \bibinfo {pages} {6280} (\bibinfo {year}
  {2018})}\BibitemShut {NoStop}%
\bibitem [{\citenamefont {Chan}(1999)}]{chan1999fresh}%
  \BibitemOpen
  \bibfield  {author} {\bibinfo {author} {\bibfnamefont {G.~K.-L.}\
  \bibnamefont {Chan}},\ }\href {https://doi.org/10.1063/1.478357} {\bibfield
  {journal} {\bibinfo  {journal} {J. Chem. Phys.}\ }\textbf {\bibinfo {volume}
  {110}},\ \bibinfo {pages} {4710} (\bibinfo {year} {1999})}\BibitemShut
  {NoStop}%
\bibitem [{\citenamefont {Cohen}\ \emph {et~al.}(2008)\citenamefont {Cohen},
  \citenamefont {Mori-S{\'a}nchez},\ and\ \citenamefont
  {Yang}}]{cohen2008fractional}%
  \BibitemOpen
  \bibfield  {author} {\bibinfo {author} {\bibfnamefont {A.~J.}\ \bibnamefont
  {Cohen}}, \bibinfo {author} {\bibfnamefont {P.}~\bibnamefont
  {Mori-S{\'a}nchez}}, \ and\ \bibinfo {author} {\bibfnamefont
  {W.}~\bibnamefont {Yang}},\ }\href
  {https://doi.org/10.1103/PhysRevB.77.115123} {\bibfield  {journal} {\bibinfo
  {journal} {Phys. Rev. B}\ }\textbf {\bibinfo {volume} {77}},\ \bibinfo
  {pages} {115123} (\bibinfo {year} {2008})}\BibitemShut {NoStop}%
\bibitem [{\citenamefont {Gould}\ and\ \citenamefont
  {Toulouse}(2014)}]{gould2014kohn}%
  \BibitemOpen
  \bibfield  {author} {\bibinfo {author} {\bibfnamefont {T.}~\bibnamefont
  {Gould}}\ and\ \bibinfo {author} {\bibfnamefont {J.}~\bibnamefont
  {Toulouse}},\ }\href {https://doi.org/10.1103/PhysRevA.90.050502} {\bibfield
  {journal} {\bibinfo  {journal} {Phys. Rev. A}\ }\textbf {\bibinfo {volume}
  {90}},\ \bibinfo {pages} {050502} (\bibinfo {year} {2014})}\BibitemShut
  {NoStop}%
\bibitem [{\citenamefont {Hodgson}\ \emph {et~al.}(2017)\citenamefont
  {Hodgson}, \citenamefont {Kraisler}, \citenamefont {Schild},\ and\
  \citenamefont {Gross}}]{hodgson2017interatomic}%
  \BibitemOpen
  \bibfield  {author} {\bibinfo {author} {\bibfnamefont {M.~J.}\ \bibnamefont
  {Hodgson}}, \bibinfo {author} {\bibfnamefont {E.}~\bibnamefont {Kraisler}},
  \bibinfo {author} {\bibfnamefont {A.}~\bibnamefont {Schild}}, \ and\ \bibinfo
  {author} {\bibfnamefont {E.~K.}\ \bibnamefont {Gross}},\ }\href
  {https://doi.org/10.1021/acs.jpclett.7b02615} {\bibfield  {journal} {\bibinfo
   {journal} {J. Phys. Chem. Lett.}\ }\textbf {\bibinfo {volume} {8}},\
  \bibinfo {pages} {5974} (\bibinfo {year} {2017})}\BibitemShut {NoStop}%
\bibitem [{\citenamefont {Benitez}\ and\ \citenamefont
  {Proetto}(2016)}]{benitez2016kohn}%
  \BibitemOpen
  \bibfield  {author} {\bibinfo {author} {\bibfnamefont {A.}~\bibnamefont
  {Benitez}}\ and\ \bibinfo {author} {\bibfnamefont {C.~R.}\ \bibnamefont
  {Proetto}},\ }\href {https://doi.org/10.1103/PhysRevA.94.052506} {\bibfield
  {journal} {\bibinfo  {journal} {Phys. Rev. A}\ }\textbf {\bibinfo {volume}
  {94}},\ \bibinfo {pages} {052506} (\bibinfo {year} {2016})}\BibitemShut
  {NoStop}%
\bibitem [{\citenamefont {G{\"o}rling}(2015)}]{gorling2015exchange}%
  \BibitemOpen
  \bibfield  {author} {\bibinfo {author} {\bibfnamefont {A.}~\bibnamefont
  {G{\"o}rling}},\ }\href {https://doi.org/10.1103/PhysRevB.91.245120}
  {\bibfield  {journal} {\bibinfo  {journal} {Phys. Rev. B}\ }\textbf {\bibinfo
  {volume} {91}},\ \bibinfo {pages} {245120} (\bibinfo {year}
  {2015})}\BibitemShut {NoStop}%
\bibitem [{\citenamefont {Li}\ \emph {et~al.}(2017)\citenamefont {Li},
  \citenamefont {Lu},\ and\ \citenamefont {Yang}}]{li2017extending}%
  \BibitemOpen
  \bibfield  {author} {\bibinfo {author} {\bibfnamefont {C.}~\bibnamefont
  {Li}}, \bibinfo {author} {\bibfnamefont {J.}~\bibnamefont {Lu}}, \ and\
  \bibinfo {author} {\bibfnamefont {W.}~\bibnamefont {Yang}},\ }\href
  {https://doi.org/10.1063/1.4982951} {\bibfield  {journal} {\bibinfo
  {journal} {J. Chem. Phys.}\ }\textbf {\bibinfo {volume} {146}},\ \bibinfo
  {pages} {214109} (\bibinfo {year} {2017})}\BibitemShut {NoStop}%
\bibitem [{\citenamefont {Gould}\ \emph {et~al.}(2019)\citenamefont {Gould},
  \citenamefont {Libereles},\ and\ \citenamefont {Perdew}}]{gould2019what}%
  \BibitemOpen
  \bibfield  {author} {\bibinfo {author} {\bibfnamefont {T.}~\bibnamefont
  {Gould}}, \bibinfo {author} {\bibfnamefont {B.}~\bibnamefont {Libereles}}, \
  and\ \bibinfo {author} {\bibfnamefont {J.~P.}\ \bibnamefont {Perdew}},\
  }\href {https://doi.org/10.26434/chemrxiv.9917948.v1} {\bibfield  {journal}
  {\bibinfo  {journal} {ChemRxiv:9917948.v1}\ } (\bibinfo {year}
  {2019})}\BibitemShut {NoStop}%
\bibitem [{\citenamefont {Hellgren}\ and\ \citenamefont
  {Gould}(2019)}]{hellgren2019strong}%
  \BibitemOpen
  \bibfield  {author} {\bibinfo {author} {\bibfnamefont {M.}~\bibnamefont
  {Hellgren}}\ and\ \bibinfo {author} {\bibfnamefont {T.}~\bibnamefont
  {Gould}},\ }\href {https://doi.org/10.26434/chemrxiv.8141912.v2} {\bibfield
  {journal} {\bibinfo  {journal} {ChemRxiv:8141912.v2}\ } (\bibinfo {year}
  {2019})}\BibitemShut {NoStop}%
\bibitem [{\citenamefont {Kraisler}\ and\ \citenamefont
  {Schild}(2019)}]{kraisler2019discontinuous}%
  \BibitemOpen
  \bibfield  {author} {\bibinfo {author} {\bibfnamefont {E.}~\bibnamefont
  {Kraisler}}\ and\ \bibinfo {author} {\bibfnamefont {A.}~\bibnamefont
  {Schild}},\ }\href {https://doi.org/10.26434/chemrxiv.9902582.v1} {\bibfield
  {journal} {\bibinfo  {journal} {ChemRxiv:9902582.v1}\ } (\bibinfo {year}
  {2019})}\BibitemShut {NoStop}%
\bibitem [{\citenamefont {Levy}(1995)}]{levy1995excitation}%
  \BibitemOpen
  \bibfield  {author} {\bibinfo {author} {\bibfnamefont {M.}~\bibnamefont
  {Levy}},\ }\href {https://doi.org/10.1103/PhysRevA.52.R4313} {\bibfield
  {journal} {\bibinfo  {journal} {Phys. Rev. A}\ }\textbf {\bibinfo {volume}
  {52}},\ \bibinfo {pages} {R4313} (\bibinfo {year} {1995})}\BibitemShut
  {NoStop}%
\bibitem [{\citenamefont {Gould}\ and\ \citenamefont
  {Pittalis}(2017)}]{gould2017hartree}%
  \BibitemOpen
  \bibfield  {author} {\bibinfo {author} {\bibfnamefont {T.}~\bibnamefont
  {Gould}}\ and\ \bibinfo {author} {\bibfnamefont {S.}~\bibnamefont
  {Pittalis}},\ }\href {https://doi.org/10.1103/PhysRevLett.119.243001}
  {\bibfield  {journal} {\bibinfo  {journal} {Phys. Rev. Lett.}\ }\textbf
  {\bibinfo {volume} {119}},\ \bibinfo {pages} {243001} (\bibinfo {year}
  {2017})}\BibitemShut {NoStop}%
\bibitem [{\citenamefont {Levy}\ and\ \citenamefont
  {Zahariev}(2014)}]{levy2014ground}%
  \BibitemOpen
  \bibfield  {author} {\bibinfo {author} {\bibfnamefont {M.}~\bibnamefont
  {Levy}}\ and\ \bibinfo {author} {\bibfnamefont {F.}~\bibnamefont
  {Zahariev}},\ }\href {https://doi.org/10.1103/PhysRevLett.113.113002}
  {\bibfield  {journal} {\bibinfo  {journal} {Phys. Rev. Lett.}\ }\textbf
  {\bibinfo {volume} {113}},\ \bibinfo {pages} {113002} (\bibinfo {year}
  {2014})}\BibitemShut {NoStop}%
\bibitem [{\citenamefont {Deur}\ and\ \citenamefont
  {Fromager}(2019)}]{deur2019ground}%
  \BibitemOpen
  \bibfield  {author} {\bibinfo {author} {\bibfnamefont {K.}~\bibnamefont
  {Deur}}\ and\ \bibinfo {author} {\bibfnamefont {E.}~\bibnamefont
  {Fromager}},\ }\href {https://doi.org/10.1063/1.5084312} {\bibfield
  {journal} {\bibinfo  {journal} {J. Chem. Phys.}\ }\textbf {\bibinfo {volume}
  {150}},\ \bibinfo {pages} {094106} (\bibinfo {year} {2019})}\BibitemShut
  {NoStop}%
\bibitem [{\citenamefont {Janak}(1978)}]{janak1978proof}%
  \BibitemOpen
  \bibfield  {author} {\bibinfo {author} {\bibfnamefont {J.~F.}\ \bibnamefont
  {Janak}},\ }\href {https://doi.org/10.1103/PhysRevB.18.7165} {\bibfield
  {journal} {\bibinfo  {journal} {Phys. Rev. B}\ }\textbf {\bibinfo {volume}
  {18}},\ \bibinfo {pages} {7165} (\bibinfo {year} {1978})}\BibitemShut
  {NoStop}%
\bibitem [{\citenamefont {Fuks}\ \emph {et~al.}(2013)\citenamefont {Fuks},
  \citenamefont {Farzanehpour}, \citenamefont {Tokatly}, \citenamefont {Appel},
  \citenamefont {Kurth},\ and\ \citenamefont {Rubio}}]{fuks2013time}%
  \BibitemOpen
  \bibfield  {author} {\bibinfo {author} {\bibfnamefont {J.~I.}\ \bibnamefont
  {Fuks}}, \bibinfo {author} {\bibfnamefont {M.}~\bibnamefont {Farzanehpour}},
  \bibinfo {author} {\bibfnamefont {I.~V.}\ \bibnamefont {Tokatly}}, \bibinfo
  {author} {\bibfnamefont {H.}~\bibnamefont {Appel}}, \bibinfo {author}
  {\bibfnamefont {S.}~\bibnamefont {Kurth}}, \ and\ \bibinfo {author}
  {\bibfnamefont {A.}~\bibnamefont {Rubio}},\ }\href
  {https://doi.org/10.1103/PhysRevA.88.062512} {\bibfield  {journal} {\bibinfo
  {journal} {Phys. Rev. A}\ }\textbf {\bibinfo {volume} {88}},\ \bibinfo
  {pages} {062512} (\bibinfo {year} {2013})}\BibitemShut {NoStop}%
\bibitem [{\citenamefont {Carrascal}\ \emph {et~al.}(2015)\citenamefont
  {Carrascal}, \citenamefont {Ferrer}, \citenamefont {Smith},\ and\
  \citenamefont {Burke}}]{carrascal2015hubbard}%
  \BibitemOpen
  \bibfield  {author} {\bibinfo {author} {\bibfnamefont {D.~J.}\ \bibnamefont
  {Carrascal}}, \bibinfo {author} {\bibfnamefont {J.}~\bibnamefont {Ferrer}},
  \bibinfo {author} {\bibfnamefont {J.~C.}\ \bibnamefont {Smith}}, \ and\
  \bibinfo {author} {\bibfnamefont {K.}~\bibnamefont {Burke}},\ }\href
  {http://stacks.iop.org/0953-8984/27/i=39/a=393001} {\bibfield  {journal}
  {\bibinfo  {journal} {J. Phys. Condens. Matter}\ }\textbf {\bibinfo {volume}
  {27}},\ \bibinfo {pages} {393001} (\bibinfo {year} {2015})}\BibitemShut
  {NoStop}%
\bibitem [{\citenamefont {Deur}\ \emph {et~al.}(2017)\citenamefont {Deur},
  \citenamefont {Mazouin},\ and\ \citenamefont {Fromager}}]{deur2017exact}%
  \BibitemOpen
  \bibfield  {author} {\bibinfo {author} {\bibfnamefont {K.}~\bibnamefont
  {Deur}}, \bibinfo {author} {\bibfnamefont {L.}~\bibnamefont {Mazouin}}, \
  and\ \bibinfo {author} {\bibfnamefont {E.}~\bibnamefont {Fromager}},\ }\href
  {https://doi.org/10.1103/PhysRevB.95.035120} {\bibfield  {journal} {\bibinfo
  {journal} {Phys. Rev. B}\ }\textbf {\bibinfo {volume} {95}},\ \bibinfo
  {pages} {035120} (\bibinfo {year} {2017})}\BibitemShut {NoStop}%
\bibitem [{\citenamefont {Senjean}\ \emph {et~al.}(2017)\citenamefont
  {Senjean}, \citenamefont {Tsuchiizu}, \citenamefont {Robert},\ and\
  \citenamefont {Fromager}}]{senjean2017local}%
  \BibitemOpen
  \bibfield  {author} {\bibinfo {author} {\bibfnamefont {B.}~\bibnamefont
  {Senjean}}, \bibinfo {author} {\bibfnamefont {M.}~\bibnamefont {Tsuchiizu}},
  \bibinfo {author} {\bibfnamefont {V.}~\bibnamefont {Robert}}, \ and\ \bibinfo
  {author} {\bibfnamefont {E.}~\bibnamefont {Fromager}},\ }\href
  {https://doi.org/10.1080/00268976.2016.1182224} {\bibfield  {journal}
  {\bibinfo  {journal} {Mol. Phys.}\ }\textbf {\bibinfo {volume} {115}},\
  \bibinfo {pages} {48} (\bibinfo {year} {2017})}\BibitemShut {NoStop}%
\bibitem [{\citenamefont {Deur}\ \emph {et~al.}(2018)\citenamefont {Deur},
  \citenamefont {Mazouin}, \citenamefont {Senjean},\ and\ \citenamefont
  {Fromager}}]{deur2018exploring}%
  \BibitemOpen
  \bibfield  {author} {\bibinfo {author} {\bibfnamefont {K.}~\bibnamefont
  {Deur}}, \bibinfo {author} {\bibfnamefont {L.}~\bibnamefont {Mazouin}},
  \bibinfo {author} {\bibfnamefont {B.}~\bibnamefont {Senjean}}, \ and\
  \bibinfo {author} {\bibfnamefont {E.}~\bibnamefont {Fromager}},\ }\href
  {https://doi.org/10.1140/epjb/e2018-90124-7} {\bibfield  {journal} {\bibinfo
  {journal} {Eur. Phys. J. B}\ }\textbf {\bibinfo {volume} {91}},\ \bibinfo
  {pages} {162} (\bibinfo {year} {2018})}\BibitemShut {NoStop}%
\bibitem [{\citenamefont {Carrascal}\ \emph {et~al.}(2018)\citenamefont
  {Carrascal}, \citenamefont {Ferrer}, \citenamefont {Maitra},\ and\
  \citenamefont {Burke}}]{carrascal2018linear}%
  \BibitemOpen
  \bibfield  {author} {\bibinfo {author} {\bibfnamefont {D.~J.}\ \bibnamefont
  {Carrascal}}, \bibinfo {author} {\bibfnamefont {J.}~\bibnamefont {Ferrer}},
  \bibinfo {author} {\bibfnamefont {N.}~\bibnamefont {Maitra}}, \ and\ \bibinfo
  {author} {\bibfnamefont {K.}~\bibnamefont {Burke}},\ }\href
  {https://doi.org/10.1140/epjb/e2018-90114-9} {\bibfield  {journal} {\bibinfo
  {journal} {Eur. Phys. J. B}\ }\textbf {\bibinfo {volume} {91}},\ \bibinfo
  {pages} {142} (\bibinfo {year} {2018})}\BibitemShut {NoStop}%
\bibitem [{\citenamefont {Vanzini}\ \emph {et~al.}(2018)\citenamefont
  {Vanzini}, \citenamefont {Reining},\ and\ \citenamefont
  {Gatti}}]{vanzini2018spectroscopy}%
  \BibitemOpen
  \bibfield  {author} {\bibinfo {author} {\bibfnamefont {M.}~\bibnamefont
  {Vanzini}}, \bibinfo {author} {\bibfnamefont {L.}~\bibnamefont {Reining}}, \
  and\ \bibinfo {author} {\bibfnamefont {M.}~\bibnamefont {Gatti}},\ }\href
  {https://doi.org/10.1140/epjb/e2018-90277-3} {\bibfield  {journal} {\bibinfo
  {journal} {Eur. Phys. J. B}\ }\textbf {\bibinfo {volume} {91}},\ \bibinfo
  {pages} {192} (\bibinfo {year} {2018})}\BibitemShut {NoStop}%
\bibitem [{\citenamefont {Li}\ \emph {et~al.}(2018)\citenamefont {Li},
  \citenamefont {Requist},\ and\ \citenamefont {Gross}}]{li2018density}%
  \BibitemOpen
  \bibfield  {author} {\bibinfo {author} {\bibfnamefont {C.}~\bibnamefont
  {Li}}, \bibinfo {author} {\bibfnamefont {R.}~\bibnamefont {Requist}}, \ and\
  \bibinfo {author} {\bibfnamefont {E.~K.~U.}\ \bibnamefont {Gross}},\ }\href
  {https://doi.org/10.1063/1.5011663} {\bibfield  {journal} {\bibinfo
  {journal} {J. Chem. Phys.}\ }\textbf {\bibinfo {volume} {148}},\ \bibinfo
  {pages} {084110} (\bibinfo {year} {2018})}\BibitemShut {NoStop}%
\bibitem [{\citenamefont {Kim}\ \emph {et~al.}(2013)\citenamefont {Kim},
  \citenamefont {Sim},\ and\ \citenamefont {Burke}}]{kim2013understanding}%
  \BibitemOpen
  \bibfield  {author} {\bibinfo {author} {\bibfnamefont {M.-C.}\ \bibnamefont
  {Kim}}, \bibinfo {author} {\bibfnamefont {E.}~\bibnamefont {Sim}}, \ and\
  \bibinfo {author} {\bibfnamefont {K.}~\bibnamefont {Burke}},\ }\href
  {https://doi.org/10.1103/PhysRevLett.111.073003} {\bibfield  {journal}
  {\bibinfo  {journal} {Phys. Rev. Lett.}\ }\textbf {\bibinfo {volume} {111}},\
  \bibinfo {pages} {073003} (\bibinfo {year} {2013})}\BibitemShut {NoStop}%
\bibitem [{\citenamefont {Loos}(2017)}]{loos2017exchange}%
  \BibitemOpen
  \bibfield  {author} {\bibinfo {author} {\bibfnamefont {P.-F.}\ \bibnamefont
  {Loos}},\ }\href {\doibase 10.1063/1.4978409} {\bibfield  {journal} {\bibinfo
   {journal} {J. Chem. Phys.}\ }\textbf {\bibinfo {volume} {146}},\ \bibinfo
  {pages} {114108} (\bibinfo {year} {2017})}\BibitemShut {NoStop}%
\end{thebibliography}
\end{document}